\documentclass[prb,aps,twocolumn,floatfix]{revtex4-2}
\usepackage[
  colorlinks=true,
  urlcolor=blue,
  linkcolor=blue,
  citecolor=blue,
  bookmarks=false,
  pdftitle={Weyl nodes annihilation by external fields},
  pdfauthor={Faruk Abdulla}
]{hyperref}

\usepackage{commath}
\usepackage[normalem]{ulem}

\usepackage{physics}
\usepackage{import}
\usepackage{subfiles}
\usepackage{graphicx}
\graphicspath{{FigureP/}}
\graphicspath{{FigureB/}}
\usepackage{bm}
\usepackage{amsmath}
\usepackage{amssymb}
\usepackage{mathtools}
\usepackage{amsfonts}
\usepackage[all]{xy}
\usepackage{xspace}
\usepackage{accents}
\usepackage{stmaryrd}
\usepackage{tabularx}
\usepackage{extarrows}
\usepackage{float}
\usepackage{multirow}
\usepackage{makecell}
\usepackage[capitalise]{cleveref}
\usepackage[dvipsnames]{xcolor}
\usepackage[normalem]{ulem}
\usepackage{pdfpages}
\makeatletter
\AtBeginDocument{\let\LS@rot\@undefined}
\makeatother

\newcommand{\comment}[1]{}
\newcommand\beq{\begin{equation}}
\newcommand\eeq{\end{equation}}

\begin{document}

\title{Pairwise annihilation of Weyl nodes induced by magnetic fields \\ in the Hofstadter regime}


\author{Faruk Abdulla$^{1,2,}$}
\email{farukhrim@gmail.com}
\affiliation{$^1$Harish-Chandra Research Institute, A CI of Homi Bhabha National
Institute, Chhatnag Road, Jhunsi, Prayagraj (Allahabad)  211019, India}
\affiliation{$^2$Physics Department, Technion - Israel Institute of Technology, Haifa 32000, Israel}

\begin{abstract}

Weyl semimetal, which does not require any symmetry except translation for protection, 
is a robust gapless state of quantum matters in three dimensions. 
When translation symmetry is preserved, the only way to destroy a Weyl semimetal 
state is to bring two Weyl nodes of  opposite chirality close to each other to 
annihilate pairwise. An external magnetic field can destroy a pair of Weyl nodes 
(which are separated by a momentum space distance $2k_0$) of opposite chirality, 
when the magnetic length $l_B$ becomes close to or smaller than the inverse 
separation  $1/2k_0$.  In this work, we investigate pairwise annihilation of Weyl nodes 
induced by external magnetic field which ranges all the way from small to a very large value 
in the Hofstadter regime $l_B \sim a$. We show that  this pairwise annihilation in a 
WSM featuring two Weyl nodes leads to the emergence of either a normal insulator 
or a layered Chern insulator.  In the case of a Weyl semimetal with multiple Weyl nodes, 
the potential for generating a variety of states through external magnetic fields emerges. 
Our study introduces a straightforward and intuitive representation of the pairwise 
annihilation process induced by magnetic fields, enabling accurate predictions of the
phases that may appear  after pairwise annihilation of Weyl nodes.

\end{abstract}

\maketitle

\section{Introduction}

Weyl semimetals (WSMs) \cite{Murakami_2007, Wan_Savrasov_2011, Yang_Ran_2011, 
Burkov_Balents_2011, Xu_Fang_2011, Lv_Ding_2015a, Lv_Ding_2015b, Xu_Hasan_2015a, 
Xu_Hasan_2015b, Lu_Soljacic_2015} are examples of three dimensional topological semimetals  
where nondegenerate valence and conduction bands touch at an even number of 
isolated points in the 3D Brillouin zone (BZ) called Weyl nodes (WNs). Each WN carries
a topological charge and has a definite chirality. The fact that WNs carry nontrivial 
topological charges leads to existence of  special kind of surface states called surface
Fermi arc  which joins the projections of WNs of opposite chiralities onto the 
surface Brillouin zone (SBZ). 

Weyl semimetal is a robust topological state of quantum matter. 
When spatial translation symmetry is preserved, the only way to destroy the state is 
to bring two WNs of opposite chiralities (or topological charges) close to each other 
to annihilate them pairwise \cite{Murakami_2007}. Weyl semimetals which are known 
for many exotic properties such as chiral  anomaly \cite{Nielsen_Ninomiya_1983, 
Aji_2012, Zyuzin_Burkov_2012}, negative magnetoresistance \cite{Son_Spivak_2013,
Gorbar_Miransky_2014, Burkov_2015, Li_Das_2016, Lu_Shun_2017, Das_Agarwal_2019, 
Das_Singh_2020, Fontana2021Topological},  planar Hall effect \cite{Nandy_Tewari_2017, Li_Shen_2018, Shama_Singh_2020, 
Li_Yao_2023, Wei_Weng_2023}, 
and Fermi arc  mediated quantum  oscillations and 3D quantum Hall effect \cite{Potter_Vishwanath_2014, 
Zhang_Vishwanath_2016, Moll_Analytis_2016, Wang_Xie_2017, Zhang_Xiu_2019, Li_Xie_2020, 
Chang_Xing_2021,  Ma_Sheng_2021, Chang_Sheng_2022, Zhang_Naoto_2022},  
require a  presence of external magnetic fields to exhibit  the above mentioned 
properties.  However an external magnetic field, if strong enough, can couple a 
pair of WNs of opposite chirality and can potentially annihilate them to destroy the 
WSM state. The authors in the Refs.  \cite{Kim_Park_2017, Chan_Lee_2017} 
found that   pairwise annihilation of WNs  by external magnetic field can happen when 
the inverse  magnetic length  $l^{-1}_B = \sqrt{eB/\hbar}$ becomes close to or larger than the 
momentum space  separation $2k_0$ between the two WNs of opposite chirality.  

To investigate the pairwise annihilation of WNs induced by external magnetic fields, the authors in the Refs. 
\cite{Kim_Park_2017, Chan_Lee_2017} considered a simple model of  Weyl semimetal, 
with two WNs only,  in a continuum  approximation. A  Hamiltonian with two WNs located 
at $ {\bf k}_w = (k_0, 0, 0)$  and $-{\bf k}_w$ may  be approximated  in a continuum   as 
\begin{align}\label{eq:ContinuumWsm}
H_{con}({\bf k}) = (k_0^2 - {\bf k}^2) \sigma_x + k_y  \sigma_y + k_z \sigma_z. 
\end{align}
The WNs are separated along the $k_x$ axis and the distance is $2k_0$ in momentum 
space. Working with 
such a Hamiltonian of WSMs, Refs. \cite{Kim_Park_2017, Chan_Lee_2017} found that 
a pair of Weyl nodes gets annihilated  when the strength of the  magnetic field is such that $l_B$  
becomes close to $1/2k_0$ or smaller than this value. 
Such pairwise annihilation of Weyl  nodes which causes a transition from gapless 
semimetal to  an insulator,  has been   also observed  in the experiments  by measuring 
the resistivity of  Weyl materials TaP \cite{Zhang_Jia_2017}  and TaAs 
\cite{Ramshaw_McDonald_2018}  at a high applied  magnetic fields.

Working with the low-energy continuum Hamiltonian outlined in Eq. \ref{eq:ContinuumWsm} 
for a WSM presents several limitations. Firstly, the applicability of the continuum 
Hamiltonian in Eq. \ref{eq:ContinuumWsm}, derived from the full lattice model of a WSM, 
is constrained to situations where the separation $2k_0$ between the two 
Weyl nodes is relatively small. Secondly, when dealing with strong magnetic fields in the 
regime where the magnetic length $l_B$ is comparable to the lattice constant (Hofstadter 
regime, $l_B \sim a$), any continuum approximation of the complete lattice model 
exceeds its range of relevance. The efficacy of the low energy continuum Hamiltonian in 
Eq.  \ref{eq:ContinuumWsm} for a WSM with two Weyl nodes is confined to 
the conditions: $\ell_B  \gg a$ and   $1/2k_0 \gg a$. 

The preceding  discussion underscores the increasing complexity associated with 
investigating  pairwise annihilation of WNs within a continuum model of WSMs 
featuring multiple WNs (e.g.  time reversal preserved WSMs), especially when 
multiple node separations are involved.  Another crucial constraint is that a continuum 
model cannot anticipate the subsequent state following the pairwise annihilation of 
Weyl nodes. This process may yield not only normal insulating states but also states 
with  nontrivial topological characteristics. 

Addressing the aforementioned challenges can be achieved by exploring a lattice 
model of a WSM to investigate the pairwise annihilation of WNs induced by external 
magnetic fields. The authors  referenced in \cite{Abdulla_Murthy_2022}, among other 
things, did an
insightful investigation in this direction.  In their work, they specifically examined 
a complicated  lattice model of a time-reversal broken WSM which  involves 
many parameters,  with a primary emphasis on constructing  phase diagrams in the 
presence of commensurate magnetic fields.  Their findings revealed new phases
which include layered Chern insulator (LCI), insulator 
which is trivial in the  bulk but has counter propagating surface states on certain open 
surface (I$'$), and a  coexistent phase (W2$'$) where  Chern bands and WPs  coexist 
with their own Fermi arc surface states. The emergence of these diverse phases from 
a given WSM state was not immediately apparent. 

In examining the pairwise annihilation WNs, it is essential to recognize that an 
external magnetic field, aligned with the direction of separation between two WNs 
of opposite chirality, cannot  couple these two nodes. The possibility of pairwise 
annihilation by an external field arises only when the field is not parallel to the direction 
of separation between two WNs of opposite chirality. 
To streamline the computation without sacrificing the essence of the problem, we will 
presume that the external magnetic field's direction is perpendicular to the separation 
between two Weyl nodes of opposite chirality.

In this work, first, we consider a simple model of time-reversal broken WSM with only 
one parameter $k_0$ ($2k_0$ is the separation between two WNs in the momentum 
space) to understand how the pairwise annihilation of WNs induced by external magnetic fields 
can lead to different states. We show that the pairwise annihilation in a WSM with  two 
WNs leads to either a normal  insulator (no surface states) or a layered Chern insulator.  
Then based on the concept that a pair of WNs (separated by $2k_0$) gets annihilated when 
$l_B \sim 1/2k_0$,  we develop a model independent intuitive representation of pairwise 
annihilation process  (an example  in Fig. \ref{fig:FA2pd1}) induced by external magnetic fields. 
Importantly, this intuitive picture of pairwise annihilation only requires information about 
the Weyl nodes' locations and the connectivities of Fermi arcs in the surface  BZ to 
accurately  predict  the phases which can appear after pairwise annihilation of Weyl nodes. 
We apply the intuitive picture of pairwise annihilation to  demonstrate how the states like 
LCI, I$'$ and W2$'$   can be straightforwardly obtained from a  simpler WSM state, 
without resorting to any complicated model as was considered by  
the reference \cite{Abdulla_Murthy_2022}.

Second, we consider a minimal model of time-reversal preserved WSM with four Weyl 
nodes. The minimal model with four WNs has two free parameters $k_1$ and $k_2$ 
(see Fig. \ref{fig:WN_TRP}) which provide momentum space separation between 
Weyl nodes of opposite chirality. In a WSM with four WNs, there are three distinct 
perpendicular directions in which a magnetic field can be applied to induce pairwise 
annihilation of  Weyl nodes. We meticulously construct phase diagrams for each of 
the three cases by solving the  model with thorough effort. Subsequently, we assert 
that these phase diagrams can be  easily derived from the intuitive representation 
of pairwise annihilation of WNs, requiring only minimal information about the 
WNs' locations and the connectivities  of Fermi arcs on the surface BZ.

We also touched upon pairwise annihilation of WNs  by magnetic fields in a WSM with six 
Weyl nodes. We analyze a simple case where all the WNs are located in a single plane.  
The intuitive picture of pairwise annihilation of WNs  immediately predicts emergence of 
two new  coexistence phases denoted as W2$''$ and  W4$'$  in Fig. \ref{fig:WannArc56}.

The plan of the paper is as follows: In Sec. \ref{Sec:TRB}, we conduct a thorough 
examination of  pairwise annihilation of WNs  by  external magnetic  fields in a simple 
model of WSM  with  two Weyl nodes (time-reversal  broken case). Then in 
Sec. \ref{TRPWSM},  we study the pairwise annihilation in a minimal model of  
time-reversal preserved WSM with four WNs,  which  has two free  parameters -the 
separations between WNs of opposite chirality. We  discuss our findings in 
Sec. \ref{Sec:Discussion} and summarize them in Sec.  \ref{Sec:Summary}.
In the Appendix \ref{App:TRP1}, we discuss about nature of the insulating states 
which appear after pairwise annihilation of WNs in the time-reversal preserved model. 

\begin{figure*}
\centering
\includegraphics[width=1\linewidth]{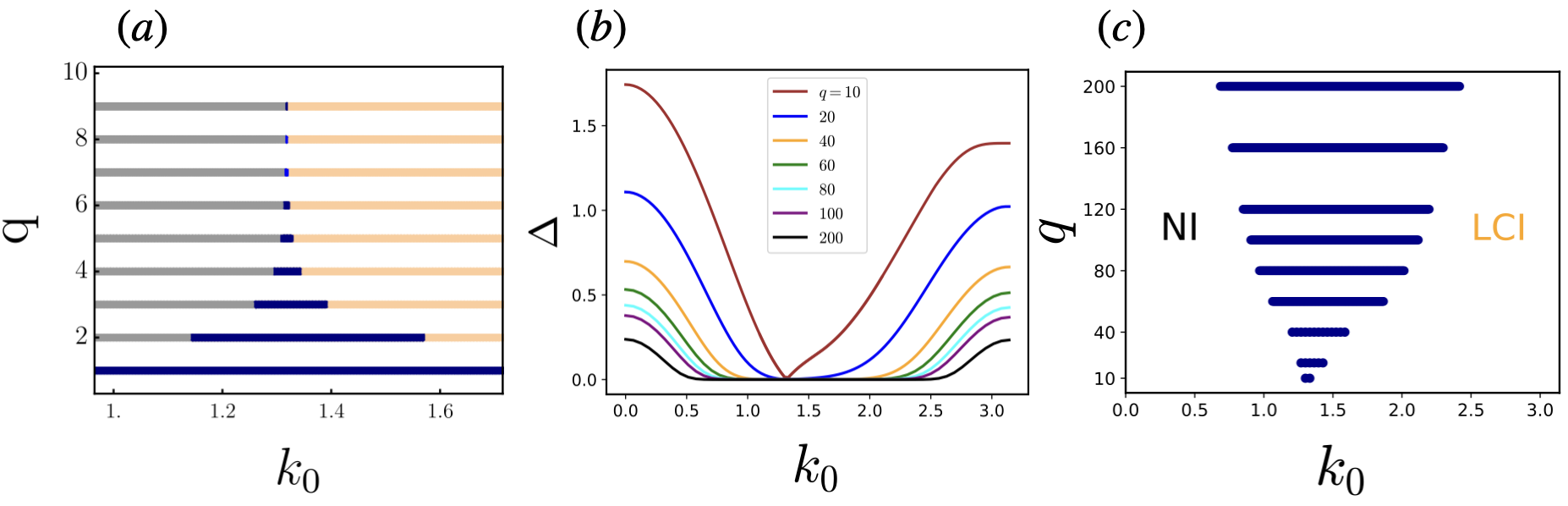}
\caption{($a$) Phase diagram of the time-reversal broken WSM (Eq. \ref{eq:H0TRB}) 
with two WNs  in presence of commensurate flux $1/q$ per unit cell, for small $q$ 
values in the Hofstadter regime $l_B \sim a$. This phase diagram is obtained from 
the gapless solution of the Bloch-Hofstadter Hamiltonian Eq. \ref{eq:blochH2}. 
The regions with grey, blue and orange color  represent a normal insulator (NI), 
WSM  and a  LCI state respectively. ($b$) Energy gap $\Delta$ is plotted as a function 
of the separation parameter $k_0$ for large $q$ values. ($c$) This phase diagram is
derived with inputs from the Fig. \ref{fig:PhaseTRB1}($b$).  We notice a similarity 
between the two phase diagrams for the small and large $q$  values (details in the text).   }
\label{fig:PhaseTRB1}
\end{figure*}


\section{Time-reversal broken WSM}
\label{Sec:TRB}

\begin{figure}[ht]
\centering
\includegraphics[width=1.0\linewidth, height=0.5\linewidth]{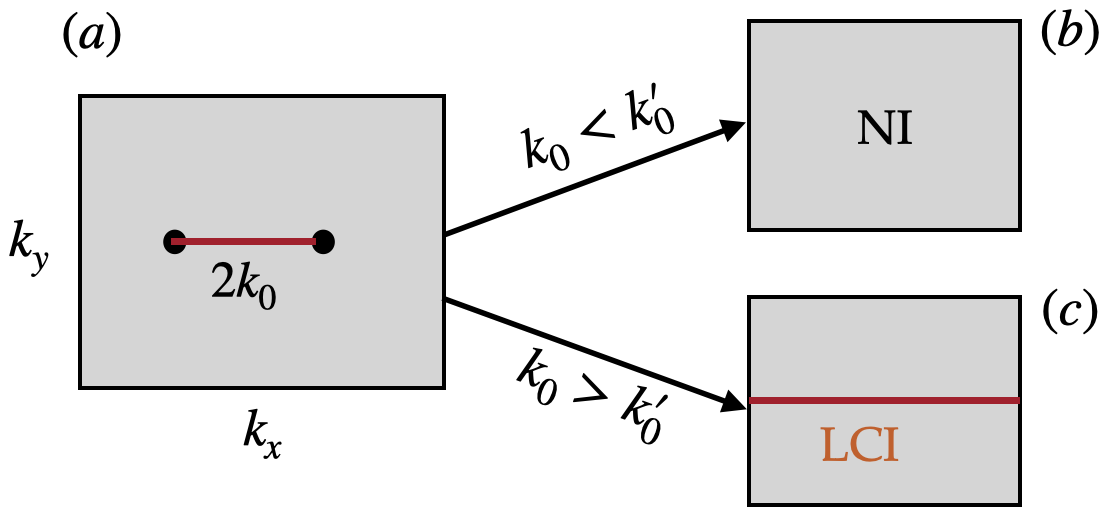} 
\caption{An intuitive picture of how and when a normal insulator (NI) and a LCI state appear 
after pairwise annihilation of two WNs separated by a momentum space distance 
$2k_0$. Figure ($a$) shows projections of the WNs (black dots) and the Fermi arc in the 
$k_x$-$k_y$ surface  BZ.  
The parameter $2k'_0 = 2\pi - 2k_0$ measure the inter-BZ  separation between
the two WNs of opposite chirality.  Two Weyl nodes get  pairwise annihilated by magnetic 
field when the inverse magnetic length $l^{-1}_B$ becomes close to or larger then 
momentum space separation between them.  There are two scenarios prevail. 
When $k_0 < k'_0$,  the inverse magnetic length $l^{-1}_B$ first reaches the intra-BZ 
separation $2k_0$. In this case, when the magnetic field is increased, two WNs approach 
each other along the Fermi arc to meet at a point inside the BZ. This leads annihilation of 
the two nodes  without leaving the  Fermi arc. Hence a normal insulator  emerges. On 
the other hand, if $k_0> k'_0$,  pairwise annihilation occurs at the boundary of the BZ by 
leaving the Fermi arc as depicted in figure ($c$). Hence  a LCI state emerges. }
\label{fig:FA2pd1}
\end{figure}

We consider the following lattice model of time-reversal broken Weyl semimetal 
\begin{equation}\label{eq:H0TRB}
\begin{aligned}
H({\bf k}) = & (2 + \cos{k_0} - \cos{k_x} - \cos{k_y} - \cos{k_z})\sigma_x \\
& + \sin{k_y}\sigma_y +  \sin{k_z}\sigma_z.   
\end{aligned}
\end{equation} 
with minimal  two WNs  at ${\bf k}_w = (k_0, 0, 0)$ and $-{\bf k}_w$ carrying monopole 
charges $C = 1$ and $-1$  respectively. The two WNs are separated along the 
$k_x$ axis by $2k_0$ and the parameter $k_0$ lies in the range $0 \le k_0 \le \pi$.  
It is easily checked that the Hamiltonian respects neither time-reversal or  particle-hole
symmetry.  However, the Hamiltonian is symmetric  under space inversion  ${\cal P} H({\bf k}) 
{\cal P}^{-1} = H(-{\bf k})$, with ${\cal P} = \sigma_x$. 
Because of the inversion symmetry, the Fermi arc which exist on  $k_x$-$k_y$
and $k_x$-$k_z$  surface BZs is a straight arc joining the projections of the 
two Weyl points (see Fig. \ref{fig:FA2pd1}$a$).  

We want to investigate pairwise annihilation of WNs by external magnetic fields for 
field's strength which ranges all the way from small  ($\ell_B \gg a$) to a
very large value ($\ell_B \sim a$) in the Hofstadter regime.  Our goal is to 
identify the states that emerge following the pairwise annihilation of Weyl nodes. 
We know for sure that the pairwise annihilation in a WSM with two WNs 
always leads to insulating states. The question we are asking is what is the nature
of these insulating  states. Specifically, we seek to determine whether the insulator 
exhibits surface states and whether it possesses topological nontriviality.

As outlined  in the 
previous section,  the applied  magnetic field  which is not aligned along  the direction 
of separation  of two WNs of opposite chirality can couple the nodes and hence can 
potentially  annihilate them.  In our  model Eq. \ref{eq:H0TRB}, the  Weyl nodes are 
separated along the  $k_x$ axis. To simplify the analysis, we assume  the magnetic field 
is aligned perpendicular to the $x$-axis, specifically aligned with the $z$-axis. 
For a WSM with two WNs separated along the $k_x$-direction, pairwise annihilation 
induced by magnetic field applied along either the $z$ or $y$-direction would result in 
identical sets of phases.

\subsection{Hofstadter Hamiltonian and gapless solutions}

An external magnetic field can be easily coupled to the  Hamiltonian in Eq. \ref{eq:H0TRB} 
by taking it to the real space, 
\begin{align}
\begin{aligned}\label{eq:RealspaceH0TRB}
H = \sum_{{\bf n}, j} c^{\dagger}({\bf n}) 2M \sigma_x c({\bf n})  -\left(c^{\dagger}({\bf n }+ 
a\hat{e}_{j}) ~ T_{j} ~ c({\bf n}) + H.c.
\right) 
\end{aligned}
\end{align}
where ${\bf n} = a(n_x, n_y, n_z)$, $n_i$ being integers, denote the lattice sites,
$\hat{e}_j$ is the unit vector along $j^{th}$  direction, and $M = 2 + \cos{k_0}$. 
The hopping matrices $T_j$, $j = (x, y, z)$, are given by: $T_x =  \sigma_x$,  
$T_y =  \sigma_x +i \sigma_y$  and $T_z =  \sigma_x + i\sigma_z$. The lattice constant 
$a$ is set to be unity for the rest of the paper.

In presence of an external magnetic fields, the hopping terms in the Hamiltonian 
Eq. \ref{eq:RealspaceH0TRB} pick up a nontrivial 
phase factor under Peierls substitution\cite{Peierls_1933}. We choose to work in the Landau 
gauge ${\bf A} = (-y, 0, 0)B$, where only the hopping in the $x$-direction picks up a nontrivial 
phase so that the Hamiltonian  in a magnetic field is obtained from Eq. \ref{eq:RealspaceH0TRB} 
by the  replacement $T_x \to  T_x \exp(-i 2 \pi y\phi/\phi_0)$. We  restrict  ourselves to 
the case where the flux $\phi$ (in units of the quantum flux $\phi_0=h/e$) per unit cell is 
commensurate i.e. $\phi/\phi_0=Ba^2/\phi_0 = p/q $, where $p$ and $q$ are relatively prime, 
so that translation symmetry along the $y$-direction is  restored with a larger unit cell \cite{Hofstadter_1976}.
In order to diagonalize the Hamiltonian, we introduce a magnetic unit cell that expands $q$
times in comparison to the original unit cell, elongating along the y-direction. Employing Fourier 
transformation in relation to the Bravais lattice positions within the magnetic unit cell yields 
following Hamiltonian
\begin{align}\label{eq:blochH2}
h_{\phi}({\bf k})  = \sum_{\alpha=0}^{q-1} & c_\alpha^{\dagger}({\bf
k})\left[f_{1}^{\alpha}({\bf k})\sigma_x + f_{3}^{\alpha}({\bf k})\sigma_z\right]
c_\alpha^{}({\bf k}) \nonumber \\
& - \left(c_{[\alpha+1]}^{\dagger}({\bf k}) e^{i q k_y \delta_{(\alpha,q-1)}} ~ T_y ~
   c_\alpha^{}({\bf k}) + H.c.\right), 
\end{align}
where  $\alpha = 0, 1, ...,q-1$  are the sublattice indices in  the magnetic unit cell 
and $\mathbf{k}$ lies in the reduced (magnetic)  Brillouin zone (MBZ),
$i.e.$, ${\bf k}$: $k_x \in \left(0, 2\pi\right)$,   $k_y \in \left(0, 2\pi/q
\right)$, $k_z \in \left(0, 2\pi\right)$. The square bracket notation in 
$c_{[\alpha+1]}^{\dagger}({\bf k})$  implies that the values of $\alpha$ are taken
modulo $q$ - $i.e.$, $[\alpha]= \alpha \mod q$.  The functions $f^\alpha_1$ and 
$f^\alpha_3$ are 
\begin{subequations}
\begin{align}
f^{\alpha}_1({\bf k}) =& 2 \left(M - \cos{\left(k_x + \frac{2\pi p}{q} \alpha\right)} -  \cos{k_z}\right)\\ 
f^{\alpha}_3({\bf k}) =& f_3({\bf k}) =  2 \sin{k_z}.
\end{align}
\end{subequations}

Note that there is only one free parameter $k_0$ (enters through $M=2 + \cos{k_0}$) 
which determines the separation  between the two WNs of opposite chirality  in our 
zero field model. Our goal is to determine the phase diagram for different values 
of $p/q$. The phase diagrams can be constructed if we can find all the gapless 
points (band touching points) of the  Hamiltonian $h_{\phi}({\bf k})$.

It is typically not possible to analytically determine the full energy spectrum of 
a Hofstadter Hamiltonian for all combinations of $p$ and $q$ values. However  
in some special cases, all the zeros (gapless solutions) of a Hofstadter  Hamiltonian  
can be found  for different values of $p$ and $q$.  Following
the Ref. \cite{Abdulla_Murthy_2022}, we find the energy spectrum of $h_{\phi}({\bf k})$  
\begin{align}
E_n({\bf k}) = \pm \sqrt{\gamma_n({\bf k}, p, q) +  \left(f_3({\bf k})\right)^2}, 
\end{align}
which is symmetric about the zero energy,  $n= 1, 2, ....,q$,  is the Landau level index and 
$\gamma_n({\bf k}, p, q) \ge 0$ for all ${\bf k}$, $p$ and $q$. 
Clearly  the zero energy  solutions are given by $ f_3({\bf k}) = \sin{k_z} = 0$ and 
$\gamma_1({\bf k}, p, q) = 0$.  The first condition tells that band touching along the 
$k_z$ direction can occur only at $k_z=0$ and/or $\pi$.  Though the quantity  
$\gamma_1({\bf k}, p, q) $ is not  known explicitly (as a function of  ${\bf k}, p$ and  
$q$),  $\gamma_1({\bf k}, p, q) = 0$  can be solved  exactly for all   ${\bf k}, p$  and  
$q$ values.  Solving  $\gamma_1({\bf k}, p, q) = 0$, we find band touching along 
the $k_y$ and $k_z$  directions can happen only at $k_y=0$ and $k_z=0$ respectively. 
We notice that the external magnetic field (aligned along the $z$-direction) did not alter 
the band touching points along the $k_y$ and $k_z$ directions i.e. the touching point
remains at $k_y=k_z=0$. 
The corresponding $k_x$  values are given by \cite{Abdulla_Murthy_2022}
\begin{align}\label{eq:GapConditionTRB}
\cos{qk_x} = (-1)^{p}\left[-T_q(g) +  2^{q-1} ~ \right],
\end{align}
where $g=1+\cos{k_0}$ and $T_q(g)$  is a Chebyshev polynomial of degree $q$ of first 
kind. The gapless solutions (exists only when the R.H.S of Eq. \ref{eq:GapConditionTRB} 
lies in the range $[-1, 1]$) describes isolated point touchings which  are the Weyl nodes 
in the theory. The WNs remain separated along the $k_x$-direction.
Note that  though the integer $p$ of  flux $p/q$ can change the sign of the R.H.S of 
Eq. \ref{eq:GapConditionTRB}, it does not affect the region of gapless solutions and 
hence the phase diagrams. In what follows we will assume $p=1$, unless it is  stated.

\begin{figure*}[t]
\centering
\includegraphics[width=0.48\linewidth, height=0.25\linewidth]{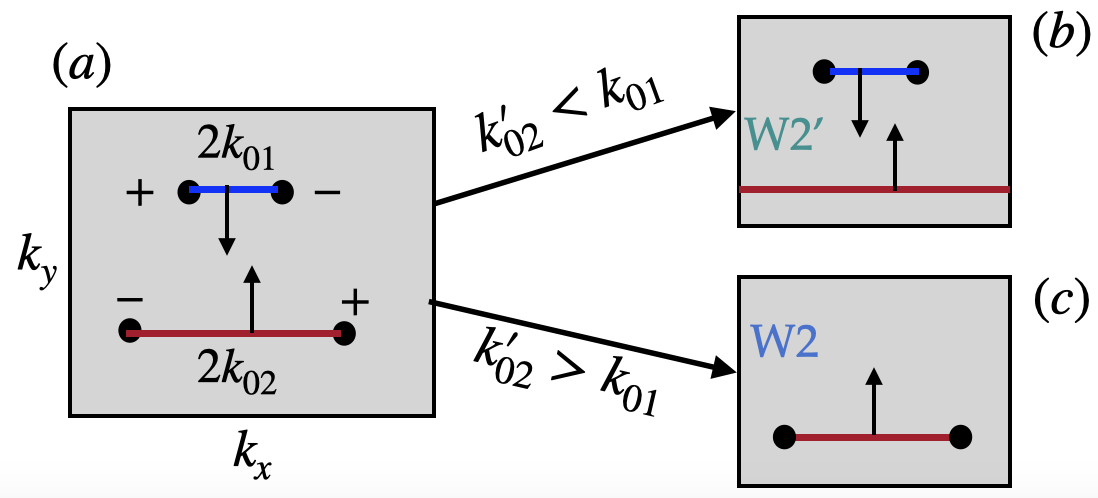} 
\includegraphics[width=0.48\linewidth, height=0.25\linewidth]{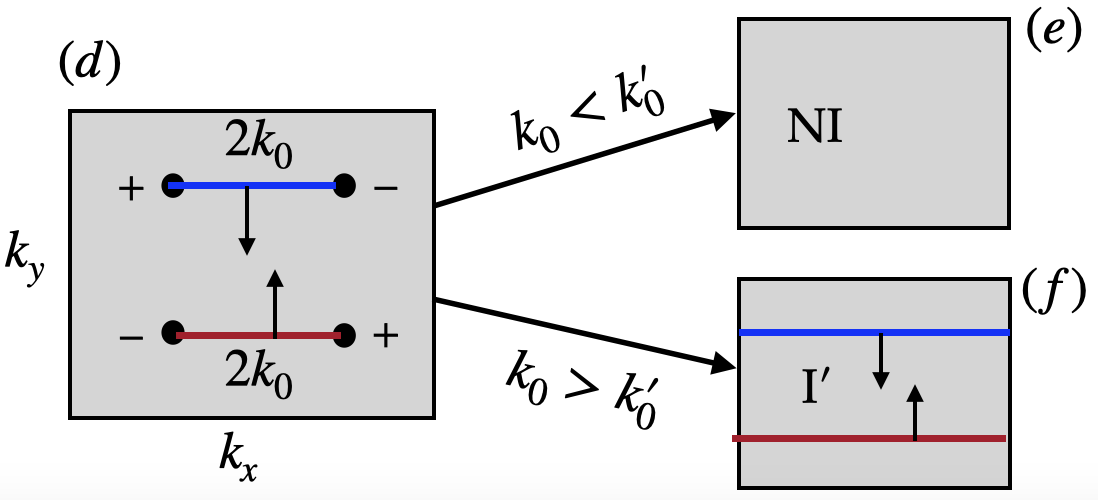} 
\caption{An intuitive representation of pairwise annihilation process of WNs of opposite 
chirality by an external magnetic field. Figures ($a$) and ($d$)  show the projections of 
the WNs (black dots) and the Fermi arcs on the  $k_x$-$k_y$ surface BZ.  For a magnetic 
field aligned in the $y$-direction, separations of WNs along the $k_x$ direction are 
relevant for pairwise annihilation. $2k_{01}$ and $2k_{02}$ are the intra-BZ separations 
and $2k'_{01} = 2\pi - 2k_{01}$ and $2k'_{02}= 2\pi - 2k_{02}$ are the corresponding 
inter-BZ separations. If $k'_{02}<k_{01}$, the pair of WNs separated by $2k_{02}$ 
will be annihilated at the boundary of BZ by leaving the Fermi arc states.  Thus a 
coexistence phase W2$'$ emerges (see figure($b$)). If  $k'_{02}>k_{01}$, then the pair of WNs 
separated by $2k_{01}$  will be annihilated  at some point inside the BZ  without 
leaving the Fermi arcs. This results in a WSM (labelled W2) with two Weyl 
nodes (see ($c$)).  Suppose $k_{01}=k_{02} = k_0$ as shown in ($d$). Now, it is clear 
that a normal insulator (NI) emerges when $k_0<k'_0$,  and an insulator (I$'$) with 
counter propagating  surface states appears   when $k_0>k'_0$.  Note that we would 
get the same set of phases if the magnetic field was aligned in the $z$-direction, provided 
the separations of the WNs along the $k_y$ direction is kept maximum.
}
\label{fig:WannArc23}
\end{figure*}

\subsection{Phase diagrams}

The phase diagram, for a given  $q$,  can be obtained from the Eq.  \ref{eq:GapConditionTRB}  
by solving it for allowed $k_0$ values such that the R.H.S remains in the range $[-1, 1]$.  
The phase diagram is shown in  Fig. \ref{fig:PhaseTRB1}a for small $q$ values. There
are two insulating regions in the phase diagram. We note that the gapless condition 
Eq.  \ref{eq:GapConditionTRB} is not enough to determine  the nature of  the two insulators 
in the phase diagram. We use numerics (to compute Chern numbers and surface states) to 
find the nature of  the two insulators  in the phase  diagram. We will shortly see that a simple 
intuitive representation of pairwise annihilation of WNs by external fields can accurately predict 
the entire phase diagram including the nature of the insulators. 

The WSM state, which existed for $k_0$ in the range $0<k_0<\pi$, now in an applied 
magnetic field exists in a smaller region (see Figs. \ref{fig:PhaseTRB1}$a$ and 
\ref{fig:PhaseTRB1}$c$). The WSM
states with either small  or large separation of WNs get gapped out first by 
the applied magnetic field and transform to insulators. 
The nature of the resulting insulators depends on the separation and the Fermi arc 
connectivity between the two  WNs in the zero field model. In our model, the Fermi arc 
is an intra-BZ  
straight  arc (along $k_x$)  joining the projections of  WNs of opposite chirality as shown 
in Fig. \ref{fig:FA2pd1}a.  We observe  that  a WSM state with small separation 
between the WNs produces a normal insulator, while a WSM state characterized by a 
large separation  between the  WNs  gives rise  to a layered  Chern insulator 
(LCI) \cite{Abdulla_Murthy_2022} after pairwise annihilation. The  LCI state carries 
nonzero Chern numbers  $C(k_x)=1$ for all $k_x$ values.

Let us take a closer look at the phase diagram Fig. \ref{fig:PhaseTRB1}$a$ for small 
$q$ values. We notice that as $q$ is decreased,  the region of the gapless WSM state 
expands. From physical point of view, this may seem counterintuitive because the 
magnetic length $l_B = \sqrt{\hbar/eB} = \sqrt{q} a$ ($a$ lattice constant) decreases with 
decreasing $q$ and hence we expect the region of the gapless WSM state in the phase 
diagram Fig. \ref{fig:PhaseTRB1}a  to contract  with decreasing $q$ (recall pairwise 
annihilation occurs when $l_B \lesssim 1/k_0$). Actually, this behavior of the system 
for small $q$ values in the Hofstadter regime $l_B \sim a$  is not contradictory but is 
consistent with what we expect in a  lattice:  The system should go towards the zero 
field limit  as we decrease 
$q$ because in the limit  $q\to 1$, the phase factor $\exp(- i 2 \pi y_j  p/q )$ ($y_j$ is 
an integer) in the hopping term  also  approaches $1$. 

Now let us focus on higher $q$ values for which the magnetic length is much 
larger than the lattice constant. From the phase diagram Fig. \ref{fig:PhaseTRB1}$a$, 
we see that the gapless region shrinks as $q$ is increased. From the gapless 
condition Eq. \ref{eq:GapConditionTRB}, we find that  the gapless region actually 
shrinks almost to a point for $q$ value as  small as   $q \sim 10 $. 
This implies that for $q \gtrsim10$ the transition from  the normal insulator to  the 
LCI state goes through  a point  instead of a region in  the $k_0$-space. So the system
remains  gapless (WSM) only at the point  $k_{0c} = r \frac{\pi}{2}$ ($r \approx 0.84$)
for $q \gtrsim10$.

This  apparently means that an applied magnetic field with very large values of 
$q$ i.e. an  arbitrarily small field can destroy a WSM  state with two WNs of  
arbitrary separation. From a physical point of view, an arbitrary small field 
cannot destroy a WSM state. There must be some additional information which is missing  
when we construct  phase diagram from the gapless condition Eq. \ref{eq:GapConditionTRB}
only. A crucial information which is missing  is that the energy gap  ($\Delta$), in the insulating 
states, falls  exponentially \cite{Abdulla_Murthy_2022}  with increasing $q$. 
Therefore  to find the correct phase diagram for  large values of $q$ in the regime 
$l_B \gg a$, we need  to compute the energy gap $\Delta$  as a function of $k_0$
and $q$. 
We have computed the energy gap $\Delta$ numerically as a function of  $k_0$ 
and plotted it for a series of values of $q$ in Fig. \ref{fig:PhaseTRB1}$b$. Now we 
find that the gapless  WSM  state exists in a finite region for a large value of $q=200$. 
We also see that  the gapless  region contracts as we decrease the value of $q$ from 
200 to 100, 80, 60,...,10, which is according to our expectation: With decreasing $q$, 
the magnetic length   $l_B = \sqrt{q} a$ decreases and hence the applied field 
annihilates a pair of WNs of higher  and higher separation. 

In summary, we find that  the analytically  obtained gapless condition produces 
correct phase diagram  for small $q$ values  in the regime $l_B \sim a$. Since energy 
gap (in the insulating states) decreases exponentially with increasing $q$, the 
gapless condition Eq. \ref{eq:GapConditionTRB} is not enough  to obtain correct 
phase diagram for large values of $q$  in the regime $l_B \gg a$.  For large  values 
of $q$,  we obtain phase diagram  by  computing the  energy gap numerically.

\subsection{An intuitive representation of pairwise annihilation process}

We have obeservd that pairwise annihilation of WNs induced by external fields in a 
WSM with two WNs  results in a normal insulator  when WNs are closed spaced. 
However, a LCI state emerges  when the separation between the two WNs  is large.  
The above result can be understood through a simple picture based on the argument 
that a pair of WNs get annihilated when the inverse magnetic length $l_B^{-1}$ becomes 
close to or larger  than  the separation $2k_0$ between the two Weyl nodes of opposite 
chirality. 
Note that in a periodic BZ, there are two separations between two WNs of 
opposite chirality located  at $(k_0, 0, 0)$ and $(-k_0, 0, 0)$: (i) intra-BZ separation 
$2k_0$ and (ii) inter-BZ separation  $2k'_0=2\pi -2k_0$. Clearly it is the shorter separation 
which determines how  the pair will be  annihilated by the applied  field. 
Now consider a WSM with $k_0<k'_0$.  In this case,  the inverse magnetic length $l_B^{-1}$ 
will first reach  $2k_0$. As magnetic field is increased, two WNs  approach each other  
along the Fermi arc to meet at a point inside the BZ and get annihilated without leaving 
the Fermi arc  (demonstrated in Fig. \ref{fig:FA2pd1}). This  results in the formation of a normal 
insulator which possesses no surface states.  On the other  hand if $k_0>k'_0$, the 
inverse magnetic length $l_B^{-1}$  will first hit  $2k'_0$  and  consequently the pair of WNs  
is expected to get annihilated at the boundary of  BZ  by  leaving  the surface Fermi arc states. 
This results in  a LCI state. The process is demonstrated  in the Fig. \ref{fig:FA2pd1}. Note
that the maximum separation occurs when  $k_0 = \pi/2$ or  $k'_0 = \pi/2$. This implies that a 
very strong field is needed to destroy  a WSM in which WNs are separated by a distance 
$2k_0 = \pi$. This is the reason why  the WSM state survives  in the central region of the phase 
diagram Fig. \ref{fig:PhaseTRB1}  in presence  of an external magnetic field. 

Here we want to point out that the region, in  which the WSM state survives, shrinks with the 
increase in the strength of the field. At certain field values (around $q \sim 10$), this region 
contracts to a singular point.  Based on the reasoning presented in the preceding paragraph, 
it is anticipated that this point is positioned at $k_0 = \pi/2$.
On the contrary, in the model we have considered, the gapless region  shrinks to the  point 
$k_{0c} = r \frac{\pi}{2}$, $r \approx 0.84$, as mentioned earlier. 
It is important to note that we do not consider this value of $k_{0c}$ to be universally 
applicable to all  WSMs; rather, it may be contingent on specific yet unknown details 
of the considered model.

\subsection{Pairwise annihilation in a WSM with multiple Weyl nodes}
\label{Subsec:Annihilation_Multiple_Nodes}

Examining pairwise annihilation becomes more challenging as the count of WNs 
rises because of a corresponding increase in the model's free parameters. 
Despite this complexity, the intuitive insights gained from studying pairwise annihilation 
in a WSM with two nodes can be readily extended to predict potential new states which 
can result in after pairwise annihilation in a WSM with multiple nodes.
To illustrate let us consider a WSM with four Weyl nodes placed in a magnetic field which 
is aligned  along the $z$-direction. For simplicity, let us assume 
all the four Weyl nodes are located in the $k_x$-$k_y$ plane at a constant $k_z = 0$, 
and  they are at a maximum separation of $\pi$ along the $k_y$ direction.  Clearly, 
maximum information about the location of the WNs are retained when they are 
projected on the $k_x$-$k_y$ surface BZ.  Projections  of the WNs with an illustrative Fermi 
arc connectivity on the $k_x$-$k_y$ surface BZ are  depicted in Figs. \ref{fig:WannArc23}$a$
and  \ref{fig:WannArc23}$d$. Since the WNs are located at the maximum separation along
the $k_y$ direction,  the relevant separation parameters  are $k_{01}$ and $k_{02}$ as shown 
in Fig. \ref{fig:WannArc23}$a$. Suppose $k_{02} > \pi/2$ and also $k_{02} \gg k_{01}$. Now 
if $k'_{02} < k_{01}$, then the magnetic length will first hit $k'_{02}$. In this situation as magnetic 
field is increased, the two WNs (separated by  $k_{02}$) will approach each other across the BZ 
to meet at the boundary of the BZ. 
This results in pairwise  annihilation of  the two WNs (separated by  $k_{02}$)  by 
leaving the  Fermi arc states. Thus we get a state with two WNs  but with an additional surface 
Fermi arc (see Fig. \ref{fig:WannArc23}$b$). This state is a  coexistent phase called W2$'$ 
which Ref. \cite{Abdulla_Murthy_2022} found in a complicated model  with many parameters. 
Now consider $k'_{02} > k_{01}$.  In this case,  the pairwise annihilation of the two WNs separated 
by  $k_{01}$  leads to a WSM state with only two Weyl  nodes as demonstrated in 
Fig. \ref{fig:WannArc23}$c$. 

Now it is clear that if $k_{01} = k_{02}$, then their pairwise annihilation by external magnetic 
fields would result either a normal insulator or an insulator (called I$'$) with counter propagating 
surface  states as shown in Fig. \ref{fig:WannArc23}$f$.

For a magnetic field along the $y$-direction, the separation parameters $k_{01}$ and 
$k_{02}$ are relevant only. In this case, the separation of the WNs along the $k_y$ direction 
is completely irrelevant for pairwise annihilation of WNs. Hence, pairwise annihilation by 
magnetic fields aligned in the $y$-direction would result in  an identical  set of phases as the 
previous case. We study this case in the Sec. \ref{subsec:TRP_By} in details.

The authors  referenced in \cite{Abdulla_Murthy_2022} explored an intricate model of a WSM
featuring eight Weyl nodes. They successfully addressed the complexities of the 
multi-parameter model and identified phases such as W2$'$ and I$'$ in the presence 
of a magnetic field. In our work, we have demonstrated how these phases could be derived 
from a simpler WSM model with only four Weyl nodes.  Crucially, our approach does not 
rely on a particular model; instead, all that is necessary is knowledge of the WNs' 
positions and the Fermi arc connectivity in the surface BZ. This enables us to precisely
predict the potential phases that may emerge in the presence of an external field.

\subsection{Effect of Zeeman energy on the phase diagram}  

So far, we have completely ignored effect of Zeeman energy on the  phase diagram. 
For a magnetic  field along the $z$-direction, the Zeeman term ($H_Z$) would be 
proportional to  the $\sigma_z$ i.e. $H_Z = E_Z \sigma_z $, where $E_Z \propto 1/q$.  
Addition of  this term  to the Hofstadter Hamiltonian  Eq. \ref{eq:blochH2}, 
will modify only the quantity  $f^{\alpha}_3 ({\bf k}) \to f^{\alpha}_3 ({\bf k}) + E_Z$. 
Therefore the band touching  along the $k_z$ direction will move from the point 
$k_{z0}=0$  to $k_{z0} = \sin^{-1}(E_Z/t_z)$ (hopping  along the $z$-direction is 
parametrized by $t_z$). This change in band touching along the $k_z$ direction 
flows to the quantity  $f^{\alpha}_1({\bf k})$ to cause a  shift in the parameter $k_0$. 
The final result is that the critical  point $k_0 = k^{*}_0$, at which a transition from 
WSM to insulator occurs, moves with the  change in  the Zeeman energy.  Of course, 
this would not give any new phase, but can move the phase boundary.

\begin{figure}[ht]
\centering
\includegraphics[width=0.7\linewidth]{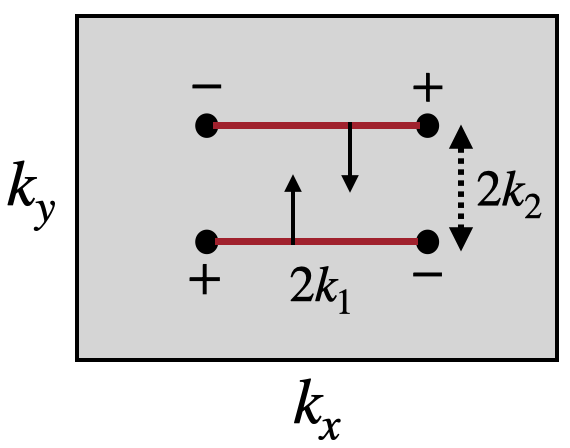}
\caption{Projections of the WNs and the Fermi arcs on the $k_x$-$k_y$ surface BZ of 
the model in Eq. \ref{eq:HKTRP}.  The arrows indicate that the states on the two Fermi arcs 
are counter propagating. Clearly,  there are two separations $2k_1$ and $2k_2$ 
between Weyl nodes  of  opposite chirality.  }
\label{fig:WN_TRP}
\end{figure}


\section{Time-reversal preserved WSM}
\label{TRPWSM}

In the previous section, we have explored pairwise annihilation in a WSM with 
two Weyl nodes. A WSM with two WNs necessarily breaks time-reversal symmetry. 
In this section, we want to examine  pairwise annihilation of WNs induced by  external 
magnetic fields in a time-reversal preserved Weyl semimetals. A minimal model 
of time-reversal preserved WSM has four Weyl nodes. Now there will be two 
independent separation parameters (see Fig. \ref{fig:WN_TRP}). In Sec. 
\ref{Subsec:Annihilation_Multiple_Nodes}, We have briefly  looked at  pairwise 
annihilation in a WSM with four WNs through the intuitive picture of pairwise 
annihilation. We restricted ourselves to a case  where separation of the WNs  
along the $k_y$  direction was fixed to simplify the  analysis. We predicted emergence 
of an insulator, labelled I$'$, with counter propagating  surface states (see Figs. 
\ref{fig:WannArc23}$d$-$f$). Here we will verify this prediction. 
Below we explore pairwise annihilation in a  WSM  with four WNs in full details. 
The separations  along both the $k_x$ and $k_y$ directions will  be considered 
as free parameters in the theory.

To study pairwise annihilation of WNs in a time-reversal preserved WSM by external 
magnetic field whose strength  can range all the way from small ($l_B \gg a$) to a very 
large value  in the  Hofstadter  regime ($l_B \sim a$),  we consider the following lattice 
model of WSM  
\begin{equation}
\begin{aligned}\label{eq:HKTRP}
H({\bf k}) = & (\cos{k_2} - \cos{k_y}) \sigma_y +  \sin{k_z}\sigma_z \\ 
& +  (1 + \cos{k_1} - \cos{k_x} - \cos{k_z})\sigma_x, 
\end{aligned}
\end{equation} 
with a minimal of four WNs located at ${\bf k}_{w_1} = (k_1, k_2, 0)$, $-{\bf k}_{w_1}$,  
${\bf k}_{w_2}  = (k_1, -k_2, 0)$,  and  $-{\bf k}_{w_2}$.  They all lie in the same 
$k_x$-$k_y$ plane at $k_z=0$. The  two  Weyl  nodes at ${\bf k}_{w_1}$ and $-{\bf k}_{w_1}$  
are time-reversal partner of each other and they carry identical chiral  charge  $\chi =1$. 
On the other hand the  pair ${\bf k}_{w_2}$  and $-{\bf k}_{w_2}$  carries  opposite chiral 
charge $\chi=-1$.  The projections of the WNs  with the Fermi arcs on the $k_x$-$k_y$ surface 
BZ are depicted  in the  Fig. \ref{fig:WN_TRP}. 

Here $\sigma$'s, which are the two by two Pauli  matrices, represent pseudo-spin degree of freedom. 
Time-reversal symmetry  ${\cal T} H({\bf k}) {\cal T}^{-1} = H({\bf -k})$ is realised by 
${\cal T} = i \sigma_x {\cal K}$,  where ${\cal K}$ acts  by taking complex 
conjugation of any  quantities appearing on the right of it.

Now we are ready to couple magnetic field to the Hamiltonian in Eq. \ref{eq:HKTRP} to 
study pairwise annihilation of WNs induced by the orbital field.  Unlike 
the time-reversal broken model with two WNs, here in the time-reversal preserved model 
with four WNs,  the external magnetic field applied along  any of the three axis direction 
can couple  the WNs and can potentially annihilate them. The minimal model 
Eq. \ref{eq:HKTRP} has two free parameters $k_1$ and $k_2$ which provide 
separations of WNs of opposite chirality as shown in the Fig. \ref{fig:WN_TRP}. 
For magnetic field applied along the $y$-direction  ($x$-direction),  the separation  
parameter $k_1$ ($k_2$) is relevant only.  This case is similar to  the two WNs'  
problem where we had  only one separation parameter. For magnetic field along 
the $y$-direction, the intuitive picture of pairwise annihilation immediately tells 
that the new state which appears after pairwise annihilation is either a normal insulator 
or the insulator I$'$ with counter propagating surface states (see Fig. 
\ref{fig:WannArc23}$f$). We will verify our prediction by solving the model for 
phase diagrams in presence of an external commensurate magnetic field. 

For magnetic  field along the $z$-direction, both the separation parameter  plays significant
role  in pairwise annihilation of  Weyl nodes. First, we  solve this model for phase diagrams 
in presence of an external commensurate magnetic field along the $z$-direction. Then, we 
argue that the phase diagrams can  be derived,  based on  the intuitive  picture of pairwise 
annihilation of WNs induced by external magnetic field.

\begin{figure*}
\centering
\includegraphics[width=0.28\linewidth, height=0.25\linewidth]{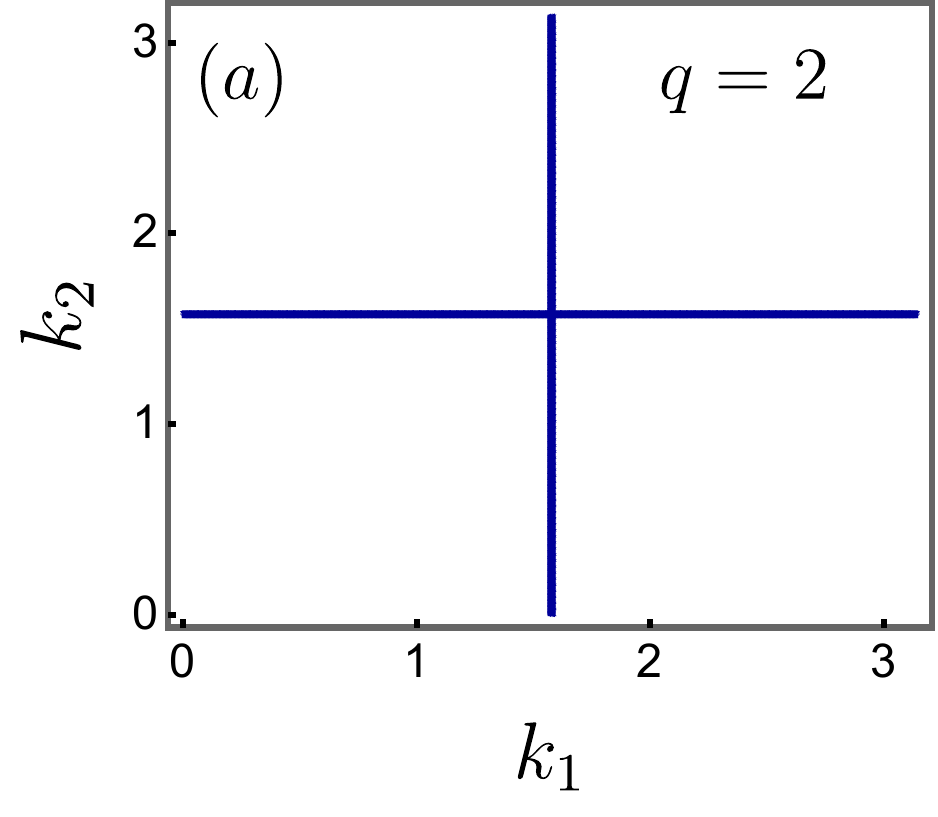}
\includegraphics[width=0.23\linewidth, height=0.25\linewidth]{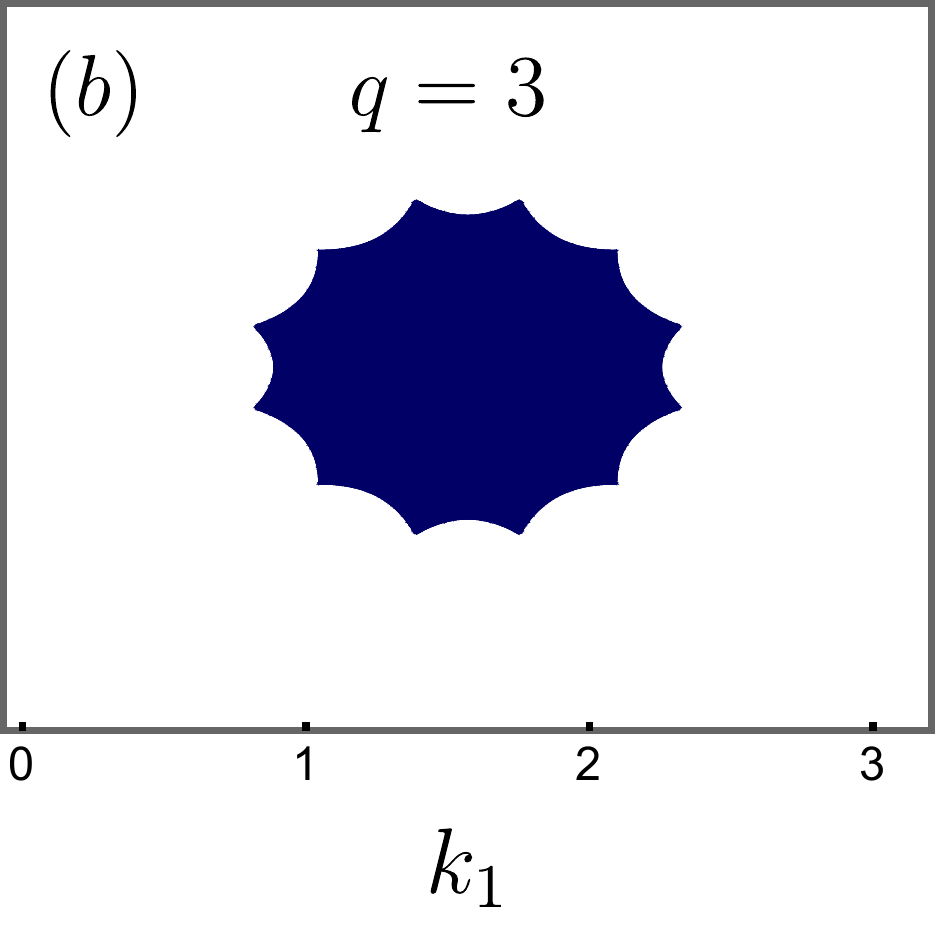}
\includegraphics[width=0.23\linewidth, height=0.25\linewidth]{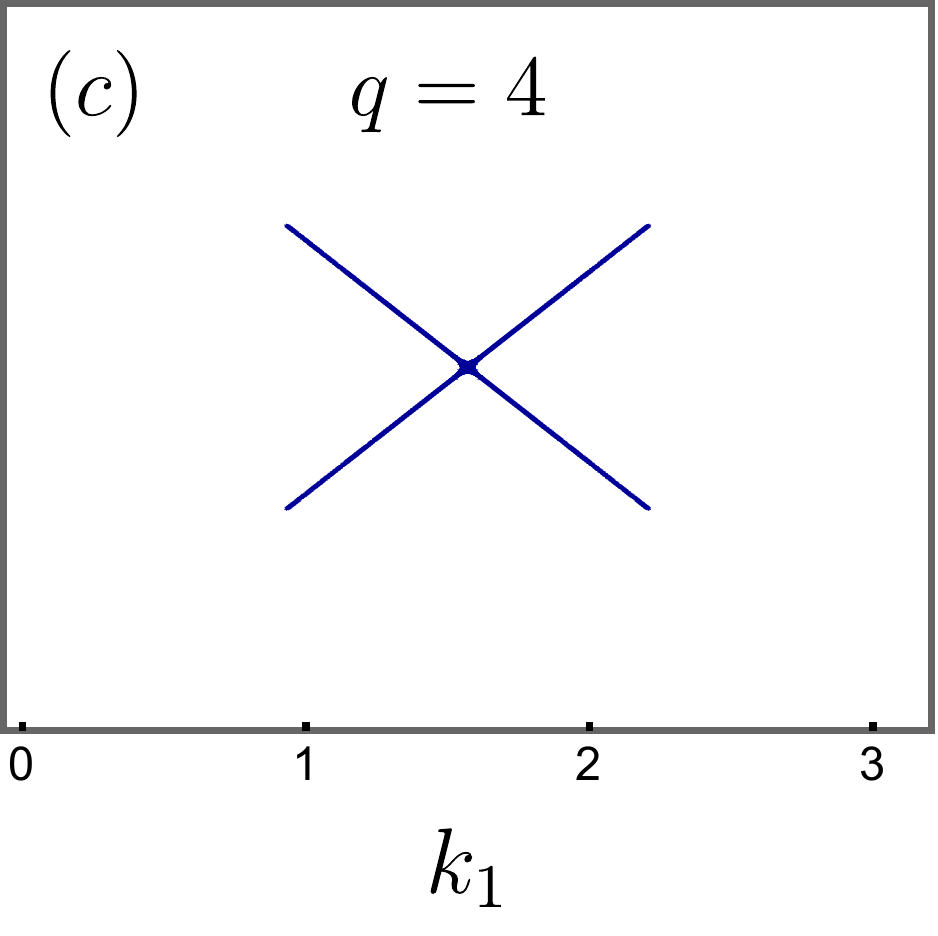}
\includegraphics[width=0.23\linewidth, height=0.25\linewidth]{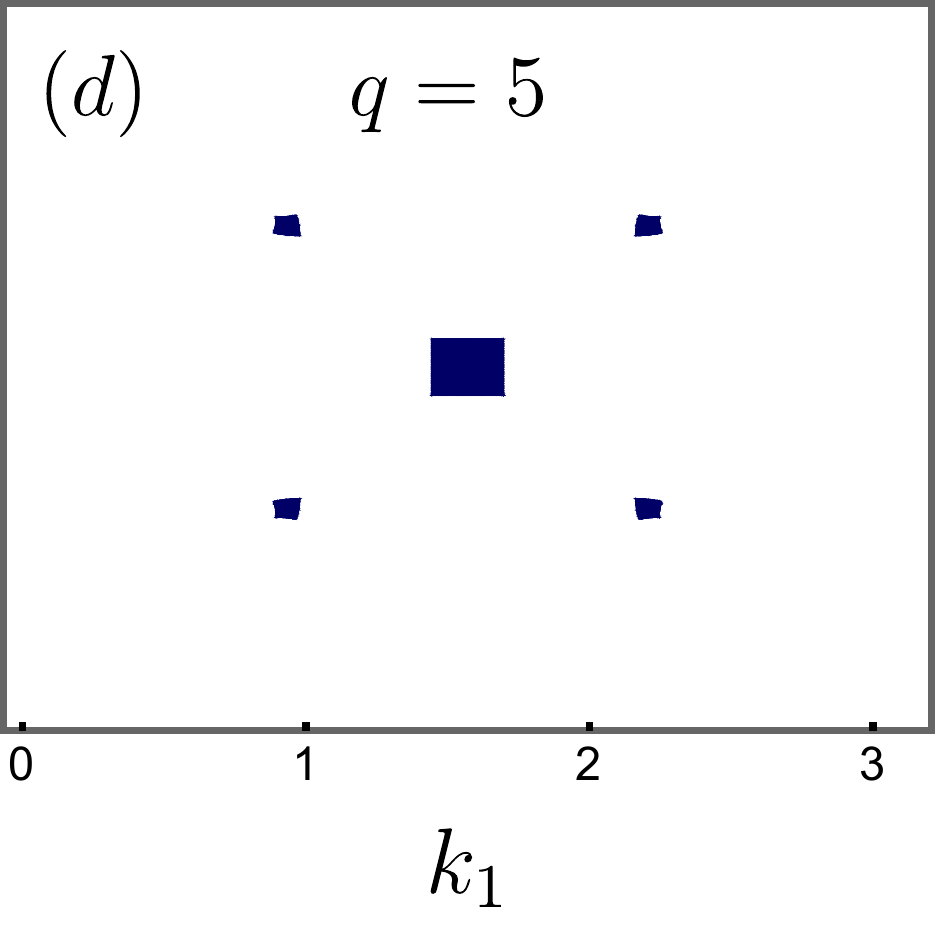}\\
\includegraphics[width=1\linewidth, height=0.5\linewidth]{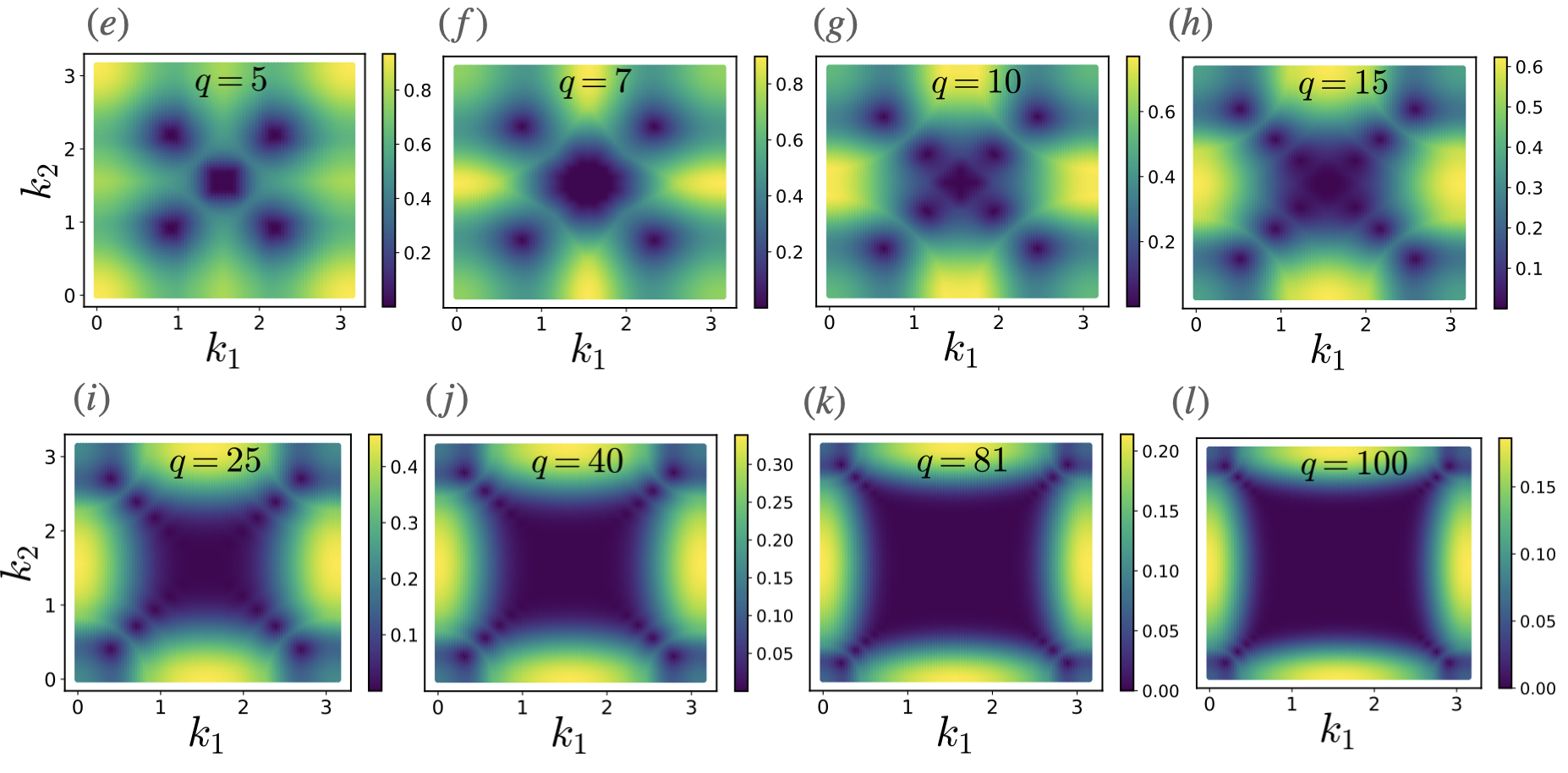}
\caption{Phase diagrams in Figs. \ref{fig:Phaseqz_TRP}($a$)-($d$) for small $q=2, 3, 4$,
and 5 are obtained from the gapless (analytical) solutions of the Bloch-Hofstadter 
Hamiltonian Eq. \ref{eq:HKTRPB}. The dark-blue region(s) represent a gapless 
phase which is  a WSM for $q=3, 5$ (odd) and nodal line semimetal for $q=2, 4$ 
(even). The white region represent a normal insulator.  For larger $q$ values, the 
phase diagrams can be derived by computing 
the energy gap as a function of the two separation parameters $k_1$ and $k_2$. 
The bulk energy gap  is computed numerically and plotted in Figs. \ref{fig:Phaseqz_TRP}
($i$-$l$) for different values of $q$. The dark-blue regions represent a gapless phase.  All
the insulating regions (in yellow) are adiabatically connected. For large $q$ value, say $q=81$,
we notice that the insulating regions appear where $|k_1 - \pi/2| \sim \pi/2$, $k_2 \sim \pi/2$
or  $k_1 \sim \pi/2$,  $|k_2 - \pi/2| \sim \pi/2$. }
\label{fig:Phaseqz_TRP}
\end{figure*}

\subsection{Field along $z$-direction}
For a constant magnetic field ${\bf B} = B \hat{z}$, we can choose the Landau  gauge 
${\bf A} = (-y, 0, 0)B$ to  work with. After going through same exercise as  in Sec. \ref{Sec:TRB},  
we arrive at the following Hofstadter  Hamiltonian 
\begin{equation}
\begin{aligned}\label{eq:HKTRPB}
h^{(z)}_{\phi}({\bf k})  = \sum_{\alpha=0}^{q-1} &  c_\alpha^{\dagger}({\bf
k})\left[f_{1}^{\alpha}({\bf k})\sigma_x +   f_{2}^{\alpha}({\bf k}) \sigma_y + f_{3}^{\alpha}({\bf k})\sigma_z\right]
c_\alpha^{}({\bf k})  \\  &  - \left(c_{[\alpha+1]}^{\dagger}({\bf k}) e^{i q k_y \delta_{(\alpha,q-1)}} ~ T_y ~
c_\alpha^{}({\bf k}) + H.c.\right) 
\end{aligned}
\end{equation}
for commensurate flux $\phi/\phi_0 = 1/q$ per unit cell. The functions $f^{\alpha}_i({\bf k})$, $i=1, 2, 3$ 
are given by 
\begin{subequations}
\begin{align}
f^{\alpha}_1({\bf k}) = & 2 \left(M - \cos{\left(k_x + \frac{2\pi p}{q} \alpha\right)} -  \cos{k_z}\right),\\ 
f^{\alpha}_2({\bf k}) \equiv & f_2({\bf k}) = 2  \cos{k_2}, \\
f^{\alpha}_3({\bf k}) \equiv & f_3({\bf k}) = 2  \sin{k_z}, 
\end{align}
\end{subequations}
where $M = 1 + \cos{k_1}$ and the hopping matrix $T_y$ in the second term of  $h^{(z)}_{\phi}({\bf k})$
is   $T_y= \sigma_y$. The  Hamiltonian $h^{(z)}_{\phi}({\bf k})$ is to be 
diagonalized in the magnetic BZ:  $k_x \in \left(0, 2\pi\right)$,   $k_y \in \left(0, 2\pi/q
\right)$, $k_z \in \left(0, 2\pi\right)$. We want to find all the gapless points in energy spectrum 
(zeros of the  Hamiltonian $h^{(z)}_{\phi}({\bf k})$) to construct the phase  diagrams for different 
values of $q$. We choose a basis $\Psi =(\psi_{\uparrow}, \psi_{\downarrow})^T$,
where $\psi_{s} = \left(c_{0,s}({\bf k}), c_{1,s}({\bf k}), ..., c_{q-1,s}({\bf k})\right)^T$ and 
$s \equiv \left(\uparrow, \downarrow\right)$, so that  $h^{(z)}_{\phi}({\bf k})$ 
can be expressed as a matrix of dimension $2q \times 2q$, 
\begin{align} \label{eq:TRPBlock1}
\Tilde{h}^{(z)}_{\phi}({\bf k}) = \begin{pmatrix}
{\bf A} & {\bf B} \\
{\bf C} & {\bf D}
\end{pmatrix}
\end{align}
where the diagonal blocks ${\bf A} = -{\bf D} = 2 \sin{k_z} {\bf I}_q$ are proportional to identity 
${\bf I}_q$ of  dimension $q\times q$,  and 
\begin{align}
{\bf B}= \begin{bmatrix}
m_0 & u & 0 & 0 & ... & u e^{ik_y q} \\
u & m_1 & u & 0 & ... & 0 \\
0 & u & m_2 & u & ... & ... \\
.. & .. & .. & .. & .. & .. \\
0 & 0 & ... & u & m_{q-2}& u \\
ue^{-ik_y q}& 0 & ... & ...& u & m_{q-1}~ \\
\end{bmatrix} = ~ {\bf C}^{\dagger}.  
\end{align}
Here $m_{\alpha} = f^{\alpha}_1({\bf k}) - i f^{\alpha}_2({\bf k})$,  $\alpha\in[0,q-1]$, and
$u=-i$. We will refer the matrix Hamiltonian $\Tilde{h}^{(z)}_{\phi}({\bf k})$ as the Bloch-Hofstadter 
Hamiltonian. The eigenvalues $E({\bf k})$ (energy spectrum of the  Hamiltonian  
$h^{(z)}_{\phi}({\bf k})$) of the the Bloch-Hofstadter Hamiltonian  are given by
\begin{align}\label{eq:det1}
\textrm{det} \begin{bmatrix}
{\bf A} - E({\bf k}) {\bf I}_q & {\bf B} \\
{\bf C} & {\bf D} - E({\bf k}) {\bf I}_q \\
\end{bmatrix} 
= 0. 
\end{align}
Since  the  diagonal blocks commutes with the off-diagonal blocks, the above 
condition reduces to
\begin{align}\label{eq:det2}
\textrm{det}\left(\gamma {\bf I}_q - {\bf B B}^{\dagger}\right) = 0, 
\end{align}
where we have used ${\bf \tilde{A} \tilde{D}} = \gamma {\bf I}_q $, $\gamma =
E^2({\bf k}) - \left(f_3({\bf k})\right)^2 $. Note that $\gamma$ is the eigenvalue of the 
positive definite matrix $ {\bf B B}^{\dagger}$, so $\gamma \ge 0$. The energy spectrum is 
\begin{align}\label{eq:EBTRP}
E_n({\bf k}) = \pm \sqrt{ \gamma_n(q, {\bf k}) + \left(f_3({\bf k})\right)^2 }, 
\end{align}
where $n=1, 2, 3, ...q$, are the Landau level indices. Clearly the spectrum is symmetric 
about the zero energy. Therefore the gapless points between the highest occupied and 
lowest unoccupied bands are given by $E_1({\bf k}) = 0$, which leads to  two separate 
conditions $f_3({\bf k}) = 2 \sin{k_z} = 0$ and $ \gamma_1(q, {\bf k}) = 0$.  The first 
condition tells that band touching along the $k_z$ direction can occur only at 
$k_{z0}=0$ and/or $\pi$. Band touchings along the $k_x$ and $k_y$ directions can be
 found from Eq. \ref{eq:det2} by setting  $\gamma = 0$. Then the Eq. \ref{eq:det2} reduces to
\begin{align}\label{eq:det3}
\textrm{det}({\bf B})  = 0, 
\end{align}
which is to be solved for a fixed $q$ to find the $k_x$ and $k_y$ values at which 
band touching can occur. Analyzing the condition in Eq. \ref{eq:det3}, we find that 
$k_{z0}=\pi$ is not an allowed solution. Therefore band touching, if occurs in presence 
of magnetic field,  along the  $k_z$ remains at $k_{z0}=0$. 
For the case of  the Hofstadter Hamiltonian in 
Eq. \ref{eq:blochH2}, a special form of the  matrix $T_y= \sigma_x + i \sigma_y$ 
brought  ${\bf B}$ in (almost) triangular form which made us possible to solve the 
above equation for arbitrary values of $q$. This is  not the situation  for the present 
case. Nevertheless  we can make a progress for small $q$ values, where the 
Eq. \ref{eq:det3} can be solved by brute force.  We have learned in 
Sec. \ref{Sec:TRB} that  the exact solution for  the zeros of the Bloch-Hofstadter 
Hamiltonian  produces correct phase diagram  in the Hofstadter regime $l_B \sim a$ 
(i.e. small $q$ values) only. For large $q$ values  in the regime $l_B\gg a$, we  
construct the phase diagrams numerically by  computing the energy gap as a function of 
the two parameters $k_1$ and $k_2$. Below  we analytically compute the zeros of the 
Bloch-Hofstadter  Hamiltonian  to  construct phase diagrams for small $q$ values 
$q=2, 3, 4$  and $5$ only.

\subsubsection{$q=2$}
Since the magnetic field is aligned along the $z$-direction, band touching point
along  the $k_z$ direction remains  at $k_{z0}= 0$  for all $q$ values. The 
corresponding $k_x$ and $k_y$ values for $q=2$ are given by the condition 
\begin{subequations}
\begin{align}
\textrm{det} \begin{bmatrix}
m_0 & u(1 + e^{-i q k_y}) \\
u(1+ e^{i q k_y})& m_1 \\
\end{bmatrix}  &  = 0 \\
m_0 m_1 - 2u^2(1 + \cos{qk_y}) &= 0, 
\end{align}
\end{subequations}
which can be simplified to a set of two conditions 
\begin{subequations}
\begin{align}
\cos{qk_x} - \cos{qk_y} &= 2(\cos^2{k_1} - \cos^2{k_2}), \\
\cos{k_1}\cos{k_2}  &= 0, 
\end{align}
\label{eq:q2gap}
\end{subequations}
We notice  that the momenta $k_x$ and $k_y$  appear only in the first of the  two 
conditions above. Therefore  the gapless solution (if exists for some $k_1$ and $k_2$) 
describes a nodal line semimetal.  The nodal line is located  in the plane  $k_z=0$.  
The full gapless  solution is shown as a shaded region in Fig. \ref{fig:Phaseqz_TRP}$a$. 
The nodal line semimetal is not a stable phase. A small change in the parameters $k_1$ 
and $k_2$ immediately gaps out the state. 

\subsubsection{$q=3$}
Solving $\textrm{det}({\bf B})=0$ for $q=3$, we get the following two conditions 
\begin{subequations}
\begin{align}
\cos{qk_x} &= F_3(\cos{k_1}, \cos{k_2}),  \\
\cos{qk_y} &= F_3(\cos{k_2}, \cos{k_1}), 
\end{align}
\label{eq:q3gap}
\end{subequations}
which $k_x$ and $k_y$ must satisfy  in order to have gapless solution.  The function 
$F_3(u, v) = 12 u v^2 - 4u^3$. Recall that bands 
touching along the $k_z$ direction  can occur only at $k_{z0}=0$.  Therefore  bands 
touching happens only at $k_{z0}=0$ and the corresponding $k_x$,  $k_y$ values 
are  determined  by Eqs. \ref{eq:q3gap}a and  \ref{eq:q3gap}b. Clearly the 
solution space  describes  point touchings which are the Weyl points in 
the theory.  A  gapless solution exists  in a  finite region in the $k_1$-$k_2$ space 
as shown  in Fig.  \ref{fig:Phaseqz_TRP}$b$. The full phase diagram  consists of only 
two phases: a topologically trivial insulating state and a gapless phase which is a WSM.

\subsubsection{$q=4$}
Solving $\textrm{det}({\bf B})=0$ for $q=4$, we get the following two conditions 
for bands touching 
\begin{subequations}
\begin{align}
& \cos{qk_x} + \cos{qk_y} = 8(\cos^4{k_1} - 6\cos{k_1}^2 \cos^2{k_2} + \cos^4{k_2}) + 2, \\
& \cos{k_1}\cos{k_2} (\cos^2{k_1} - \cos^2{k_2}) = 0. 
\end{align}
\label{eq:q4gap}
\end{subequations}
We notice that the  $k_x$,  $k_y$ values, at which bands touching can occur, are  
solely determined by the first condition Eq. \ref{eq:q4gap}a (provided the second 
condition is satisfied). The 
second condition, which involves only the two parameters $k_1$ and $k_2$ but 
no momenta, forces the gap closing to occur only on a contour (not a region) in 
the $k_1$, $k_2$  parameters space. Since the $k_x$,  $k_y$ values for bands 
touching are determined by only an one condition, the gapless solution describes 
nodal line semimetal. This is similar to what we have seen for  the case of $q=2$. 
The full phase diagram (depicted in Fig. \ref{fig:Phaseqz_TRP}$c$) consists of only 
two phases:  a topologically trivial insulator and a gapless state which is a nodal 
line semimetal. 

\subsubsection{$q=5$}
Solving $\textrm{det}({\bf B})=0$ for $q=5$, we get the following two conditions 
for a gapless solution 
\begin{subequations}
\begin{align}
\cos{qk_x} &= F_5(\cos{k_1}, \cos{k_2}),  \\
\cos{qk_y} &= F_5(-\cos{k_2}, \cos{k_1}),
\end{align}
\label{eq:q5gap}
\end{subequations}
where $ F_5(u, v) = 16u (u^4 - 10u^2 v^2 + 5v^4) + 5u(1+\sqrt{5})/2$.
Similar to  the case of $q=3$,  bands touching  for $q=5$ occurs at isolated 
points in the BZ: $k_z=0$,  and $k_x$, $k_y$ values are given by the simultaneous 
solution of the Eqs. \ref{eq:q5gap}a  and \ref{eq:q5gap}b. The band touching points 
are the Weyl points in the 
theory.  Gapless solution exists  in a  finite region in the $k_1$-$k_2$ space as shown  in Fig. 
\ref{fig:Phaseqz_TRP}$d$. Like the phase diagram for $q=3$, the phase diagram for $q=5$ 
(see Fig. \ref{fig:Phaseqz_TRP}$d$) also consists of a WSM phase  and a topologically trivial 
insulating phase only.

\begin{figure}[t]
\centering
\includegraphics[width=0.32\linewidth, height=0.5\linewidth]{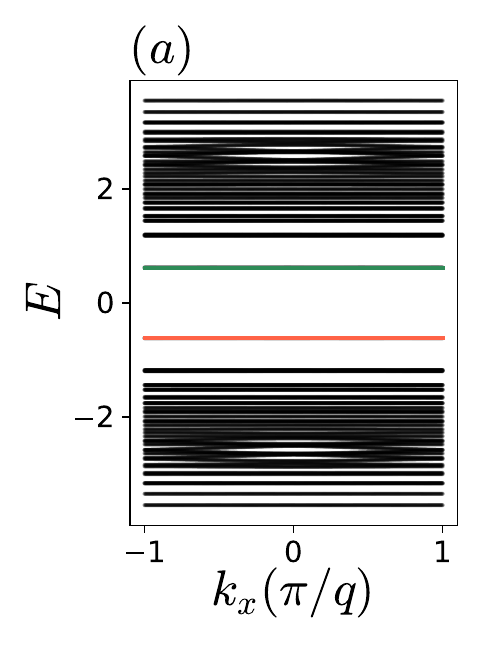}
\includegraphics[width=0.32\linewidth, height=0.5\linewidth]{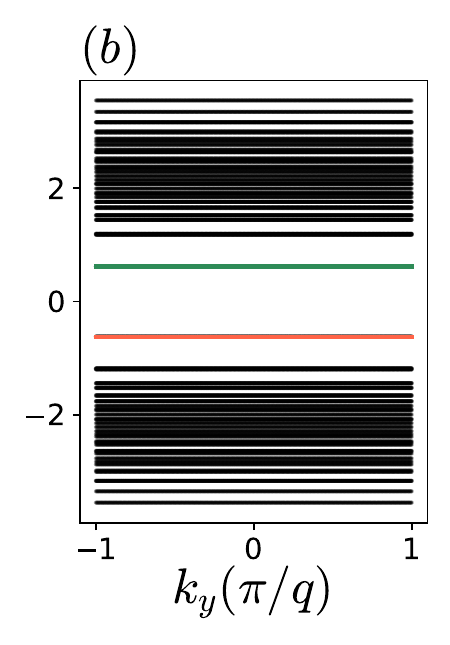}
\includegraphics[width=0.32\linewidth, height=0.5\linewidth]{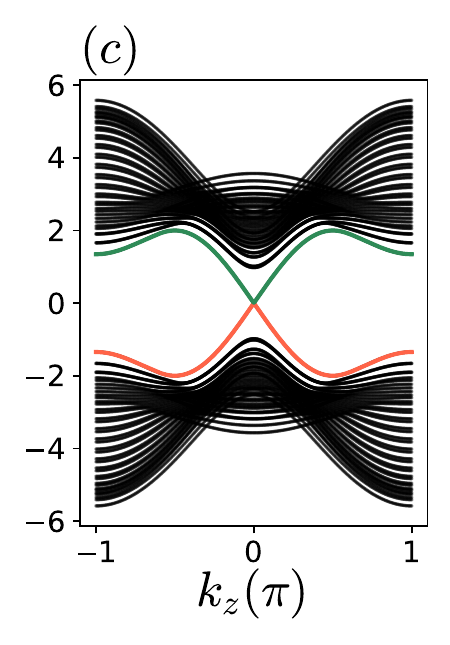}
\caption{Bulk energy dispersion  of the Hofstadter Hamiltonian $h^{(z)}_{\phi}$, 
Eq. \ref{eq:HKTRPB},  for  $q=40$. The separation parameters are $k_1=2.0$ and 
$k_2=1.4$.  For each of the plots, one of the momenta is allowed to vary, and the other 
ones are fixed  at $k_x = 0.3 \pi/q$,  $k_y = 0.2 \pi/q$ and $k_z = 0.1 \pi$ appropriately. 
Energy bands along the $k_x$ and $k_y$ directions form flat Landau levels.}
\label{fig:Enegry_q40}
\end{figure}

Finding gapless solution analytically becomes challenging  as $q$ increases. 
For $q>5$, the phase digram can be understood by computing  the energy gap as a function 
of $k_1$ and $k_2$. The result is shown in the second and third row of Fig. \ref{fig:Phaseqz_TRP}. 
We find that  every insulating region is adiabatically connected,  and all  gapless regions 
characterize  the same phase. For any odd values of $q=1, 3, 5, 7, 9, ...$, the gapless regions describe 
a Weyl semimetal state. Though for small and even values of $q=2, 4, 6,...$, the gapless regions 
describe a nodal line semimetal state, for large values of $q$, the system,  in the gapless regions,
behaves like  a WSM  in terms of the low energy dispersion.  When $q$ is significantly large 
($l_B \gg a$), it becomes challenging to differentiate the low energy spectra between even and 
odd values of $q$. As $q$ increases, the energy 
bands along the $k_x$, $k_y$ directions becomes flatter \cite{Abdulla_Murthy_2022}.  For large $q$ 
values in the regime $l_B \gg a$ (semiclassical regime), the bands along the $k_x$, $k_y$ 
directions become almost flat to form dispersionless Landau levels as expected from 
the continuum approximation in the semiclassical regime \cite{Goerbig_Piechon_2008, 
Burkov_Balents_2011b, Tchoumakov_Goerbig_2016, Li_Das_2016}. For an illustration, energy 
dispersion, for $q=40$, along all the three $k_x$, $k_y$, $k_z$ directions are depicted 
in Fig. \ref{fig:Enegry_q40}.

From the phase diagrams for large $q$ values, we observe that the WSM state gets 
gapped out in some specific regions  in the $k_1$-$k_2$ parameter space  and the
area of the insulating region increases  with the strength of the applied magnetic  
field $|{\bf B}| \propto 1/q$. Let us closely examine the phase diagram for $q = 100$
(Fig. \ref{fig:Phaseqz_TRP}$l$), and focus on how the phase diagrams evolve as the 
value of $q$ decreases.. We notice that regions, where
$|k_1 - \pi/2| \sim \pi/2$ and $k_2 \sim \pi/2$   are gapped.  Similarly, the regions where 
$k_1 \sim \pi/2$ and $|k_2 - \pi/2| \sim \pi/2$ are  also gapped. However, the region in 
which  $k_1 \sim \pi/2$ and $k_2 \sim \pi/2$ remains
gapless (WSM). As $q$ decreases (see Figs. \ref{fig:Phaseqz_TRP}$i$-$j$), the 
areas of the gapped insulating regions increase and simultaneously  the areas of the 
gapless regions decrease. All these  can be understood from the very fundamental concept 
that a pair of WNs of opposite chirality, which are separated by a momentum space 
distance $2k_0$,  annihilates each other when the magnetic  length $l_B = \sqrt{q} a$ hits
the inverse separation $1/2k_0$.  Clearly the regions with either small $k_1$   or small 
$k_2$ values will be gapped out first after pairwise annihilation of Weyl nodes. Recall
when $k_1 > \pi/2$ or $k_2 > \pi/2$,  one should compare the momentum 
space distances $k'_1 = \pi -k_1$ and  $k'_2= \pi -k_2$ with the inverse magnetic 
length. Therefore the regions with  either $|k_1 - \pi/2| \sim \pi/2$  or $|k_2 - \pi/2| \sim \pi/2$ 
will be gapped out first. The separations between WNs of opposite chirality are maximum  
in the central region $k_1 \sim \pi/2$ and $k_2 \sim \pi/2$. This is the reason why it requires 
a very strong fields to gap out the central region. 

\vspace{0.5cm}

\begin{figure*}
\centering
\includegraphics[width=0.28\linewidth, height=0.25\linewidth]{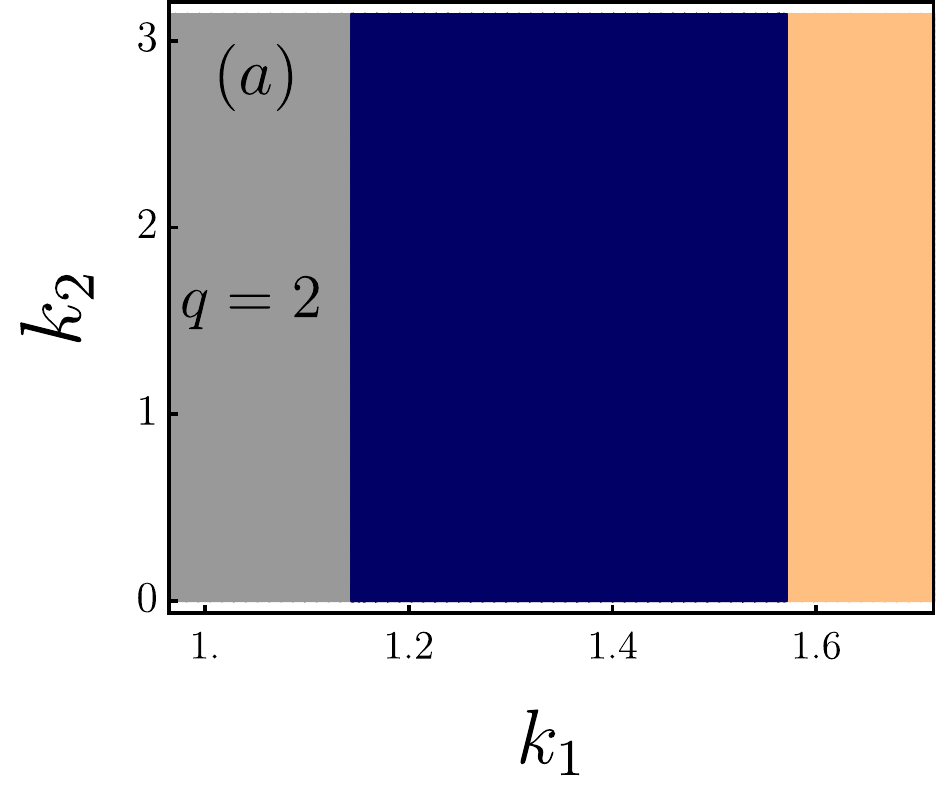}
\includegraphics[width=0.23\linewidth, height=0.25\linewidth]{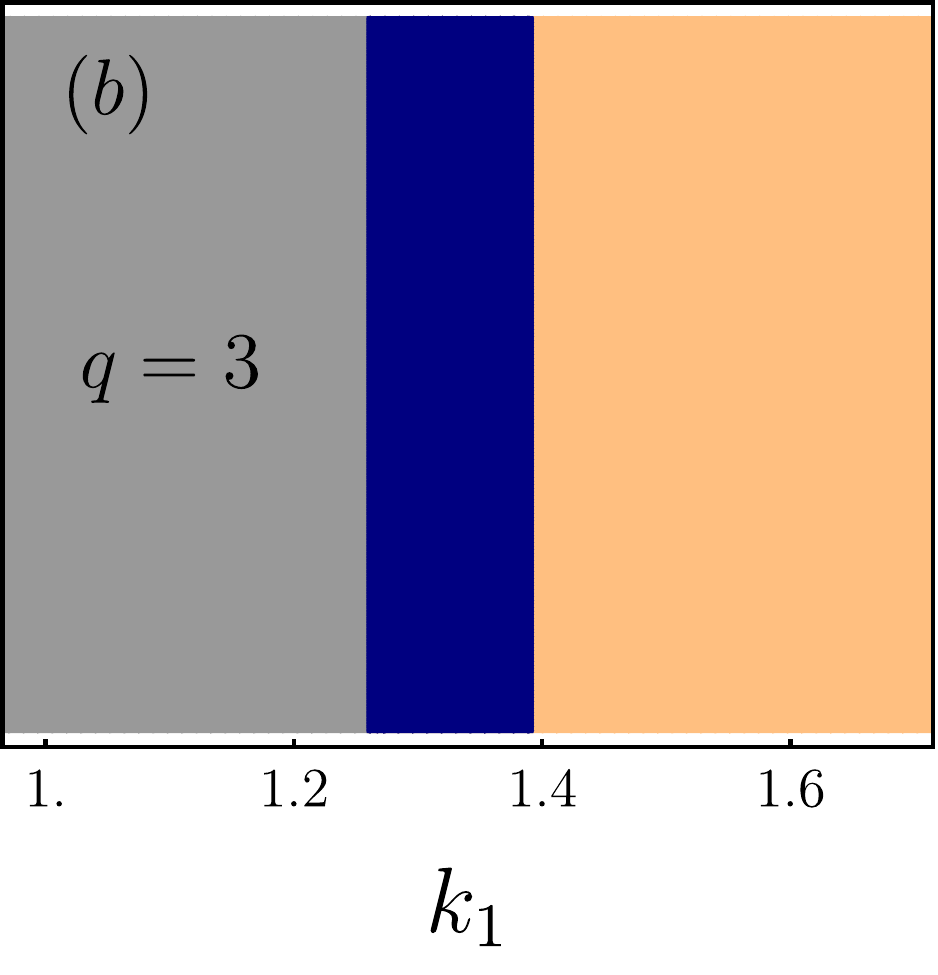}
\includegraphics[width=0.23\linewidth, height=0.25\linewidth]{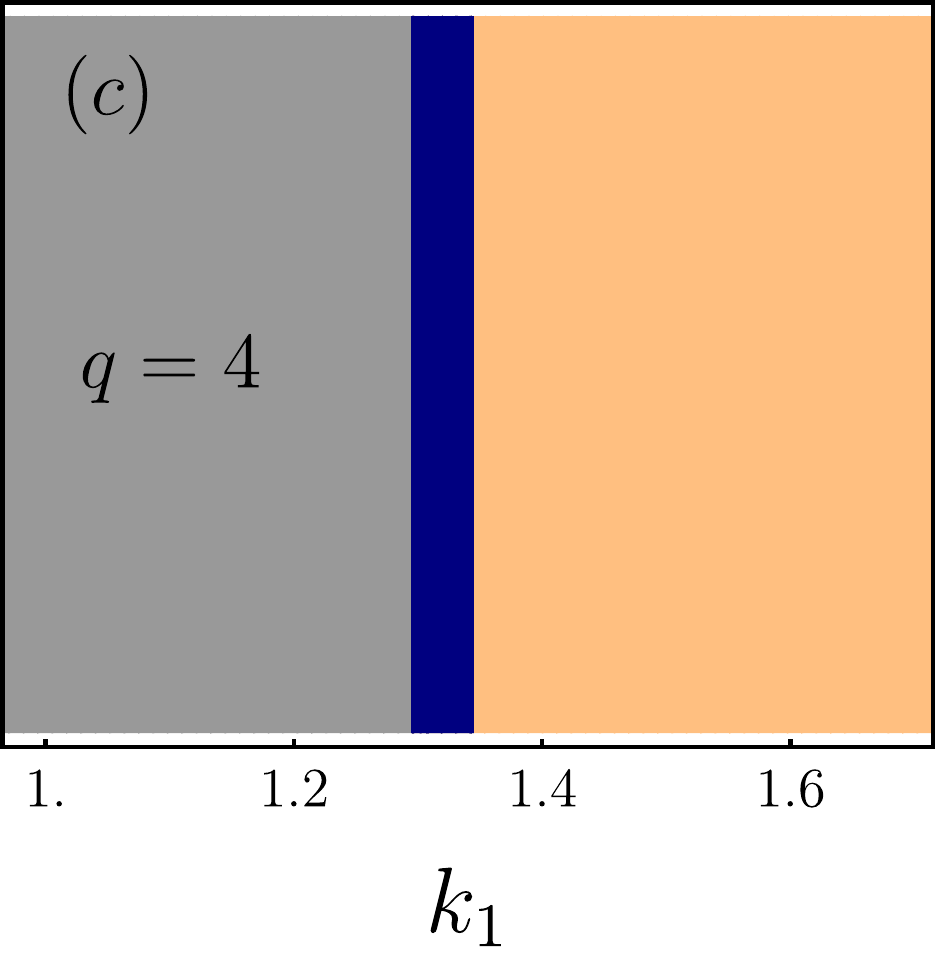}
\includegraphics[width=0.23\linewidth, height=0.25\linewidth]{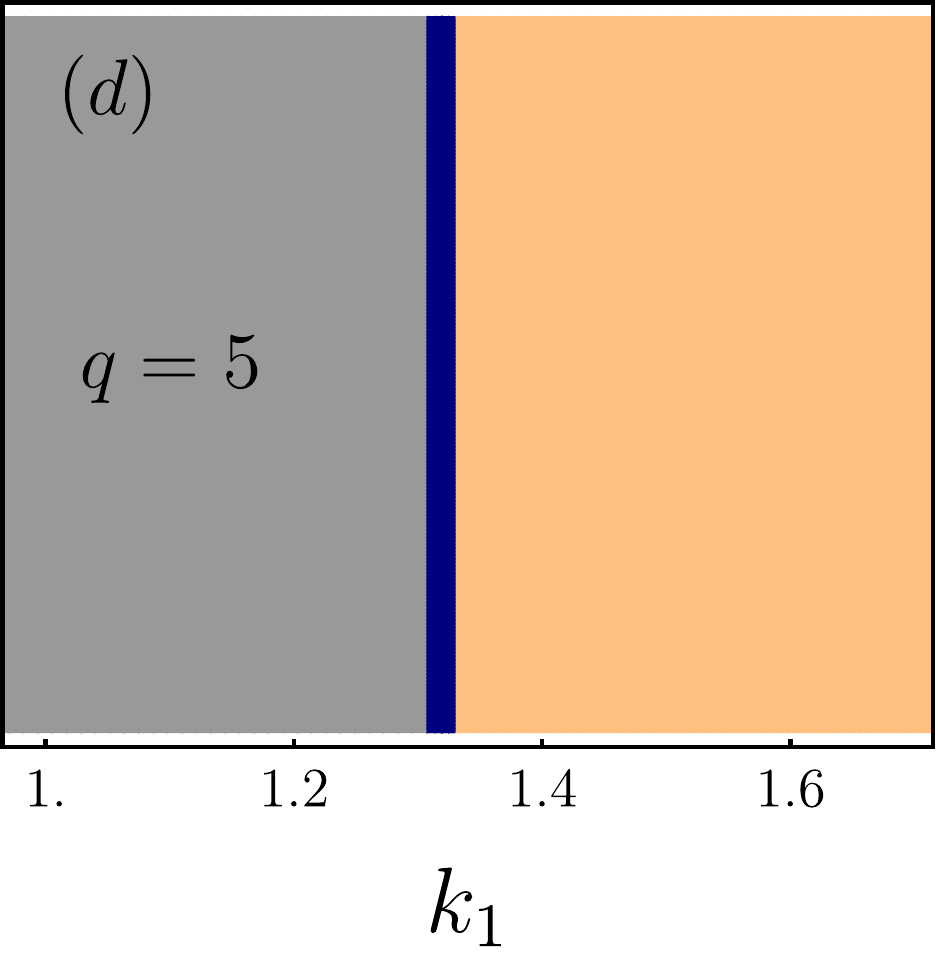} 
\caption{Phase diagrams of the time-reversal preserved WSM (Eq. \ref{eq:HKTRP}) 
with four WNs  in presence of $1/q$ commensurate flux per unit cell, along the 
$y$-direction. The shaded areas in grey, blue, and orange signify a normal insulator, 
Weyl semimetal, and an insulator (I$'$), respectively. The latter exhibits counter-propagating 
surface states along the open surface in the z-direction.}
\label{fig:Phaseqy_TRP}
\end{figure*}

An interesting tension occurs when $k_2 \approx k_1$ or $k_2 \approx \pi - k_1$. In this 
situation a single WN of chirality $\chi$ gets  simultaneously  coupled with two WNs  of 
chirality $-\chi$. This leads to an effective coupling between two WNs of same chirality. 
Since a WN cannot  annihilate another WN of same chirality, we get a partial annihilation 
of WNs along the  line  $k_2 \approx k_1$ or $k_2 \approx \pi - k_1$  in the phase diagrams.

What is common among all the phase diagrams is that there are only two phases: an insulator 
and a gapless state. Let us focus on a  phase diagram for a particular value of $q=40$. 
The entire insulating region in the phase diagram may be split into four subregions: left, 
right, top and bottom insulating regions. From the intuitive picture   of pairwise annihilation 
of WNs, we expect the insulators which are living in the   left, top  and bottom regions will 
not have any surface states. The reason is the following. In the left region where $k_1\ll \pi/2$ 
and $k_2 \sim \pi/2$,  the separation  parameter $k_1$ is relevant for pairwise annihilation. 
In this case, WNs which are separated by $2k_1$ get pairwise annihilated at some point 
inside the BZ. Hence no Fermi arc states are left.  In the bottom region ($k_2 \ll \pi/2$) and 
top region ($k_2 \sim \pi$), the separation parameter $k_2$ is relevant for pairwise annihilation. 
Since the Fermi arcs are counter propagating, pairwise annihilation of WNs either at a point  inside 
the BZ or at  the boundary of the BZ  cannot leave the Fermi arc states. We have verified this 
numerically.  However the insulator, which is  living in the right insulating region where $k'_1 \ll k_1$ 
and  $k_2 \sim \pi/2$, can have surface states 
in accordance  with our intuitive picture of pairwise annihilation of Weyl nodes (see Appendix 
\ref{App:TRP1} for details). The bulk of this insulating state is of course trivial and the state is 
adiabatically  connected to the adjacent insulating states.

\subsection{Field along $y$-direction }
\label{subsec:TRP_By}

\begin{figure}[t]
\centering
\includegraphics[width=1\linewidth, height=0.45\linewidth]{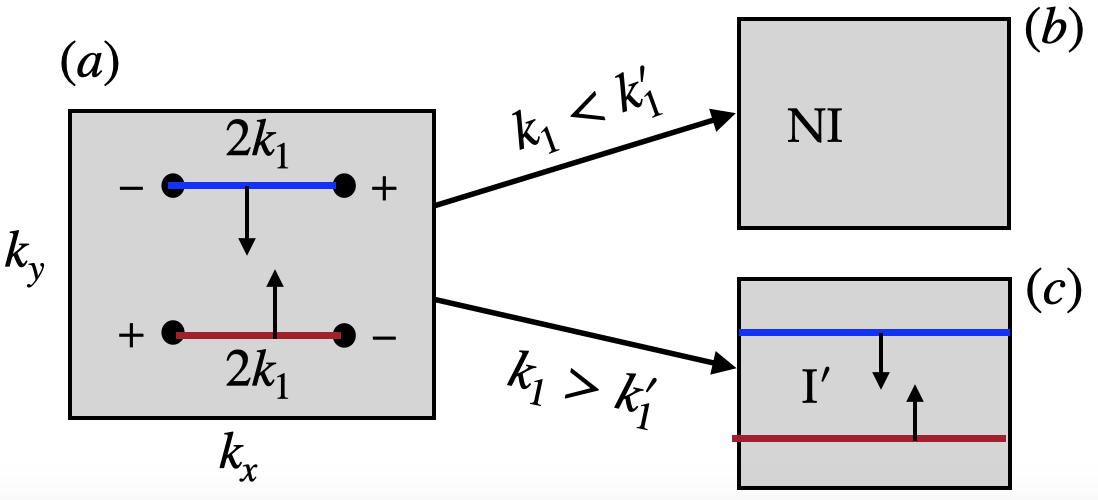} 
\caption{($a$) The Fermi arcs and the projections of the WNs  on the $k_x$-$k_y$ surface BZ
of the WSM defined in Eq. \ref{eq:HKTRP}. For magnetic field along $y$-direction, the 
separation parameter $k_1$ is relevant for  pairwise annihilation of Weyl nodes. If $k_1 < k'_1$,  
pairwise annihilation of the WMs, which  occurs at a point inside  the BZ, does not leave the 
Fermi arc  states. Hence a normal insulator results in.  If $k_1 > k'_1$,  pairwise annihilation 
occurs at the boundary of the BZ   by  leaving the  Fermi arc states. Hence, the insulator 
I$'$ emerges. }
\label{fig:WannArc4}
\end{figure}

An external magnetic field oriented in the y-direction cannot couple WNs which are separated 
along $k_y$ direction. In this case, the crucial separation to consider for pairwise annihilation 
of WNs  is $k_1$. The current problem can be thought of as a  two copies of a two 
WNs' problem,  similar to the   time-reversal broken case  studied in the Sec. \ref{Sec:TRB}. 
Here, the  separation parameter $k_1$  plays the role  of the parameter $k_0$  of the 
time-reversal broken  case (see Eq. \ref{eq:H0TRB}) with two Weyl nodes. The intuitive 
picture of pairwise annihilation  of WNs  (see Figs . \ref{fig:WannArc4} and \ref{fig:WannArc23}$d$-$f$) 
immediately tells that the phase which appear  after annihilation is either a normal insulator
or an insulator (I$'$) with counter  propagating  surface states on the $k_x$-$k_y$ surface BZ.  
In the following, we verify  this prediction by  solving the model for phase diagram in presence 
of commensurate  magnetic fields.

We choose to work with the Landau gauge ${\bf A} =(z, 0, 0)B$. In this choice of gauge, 
the Hofstadter  Hamiltonian takes  the following form (after a unitary rotation in 
$\sigma$'s space about the $x$-direction)
\begin{align}\label{eq:blochH3}
h^{(y)}_{\phi}({\bf k})  = \sum_{\alpha=0}^{q-1} & c_\alpha^{\dagger}({\bf
k})\left[f_{1}^{\alpha}({\bf k})\sigma_x + f_{3}^{\alpha}({\bf k})\sigma_z\right]
c_\alpha^{}({\bf k}) \nonumber \\
& - \left(c_{[\alpha+1]}^{\dagger}({\bf k}) e^{i q k_z \delta_{(\alpha,q-1)}} ~ T_z ~
   c_\alpha^{}({\bf k}) + H.c.\right), 
\end{align}
for commensurate flux $\phi/\phi_0 = 1/q$ per unit cell. The functions $f^{\alpha}_1({\bf k})$,
$f^{\alpha}_3({\bf k})$ are given by 
\begin{subequations}
\begin{align}
f^{\alpha}_1({\bf k}) = & 2 \left(M - \cos{\left(k_x + \frac{2\pi}{q} \alpha\right)} \right),\\ 
f^{\alpha}_3({\bf k}) \equiv & f_3({\bf k}) = 2 \left(\cos{k_2} - \cos{k_y} \right), 
\end{align}
\end{subequations}
where $M= 1+\cos{k_1}$ and the hopping matrix $T_z = \sigma_x - i \sigma_y$. Note 
that the Hofstadter Hamiltonian $h^{(y)}_{\phi}({\bf k})$ is defined in the magnetic BZ:
$k_x \in \left(0, 2\pi\right)$,   $k_y \in \left(0, 2\pi\right)$, $k_z \in \left(0, 2\pi/q \right)$.
We can obtain the phase diagrams by solving the spectrum for gapless points. Writing
$h^{(y)}_{\phi}({\bf k})$  in a matrix form (same as Eq. \ref{eq:TRPBlock1}), we obtain ${\bf A}
 = -{\bf D} = 2\left(\cos{k_2} - \cos{k_y} \right) {\bf I}_q$  and the block matrix ${\bf B}$ is
\begin{align}
{\bf B}= \begin{bmatrix}
m_0 & -2 & 0 & 0 & ... & 0 \\
0 & m_1 & -2 & 0 & ... & 0 \\
0 & 0 & m_2 & -2 & ... & ... \\
.. & .. & .. & .. & .. & .. \\
0 & 0 & ... & 0 & m_{q-2}& -2 \\
2e^{-i q k_z }& 0 & ... & ...& 0 & m_{q-1}~ \\
\end{bmatrix}, 
\end{align}
is almost an upper triangular matrix except the element $2e^{-ik_z q}$. 

\begin{figure*}
\centering
\includegraphics[width=0.48\linewidth, height=0.23\linewidth]{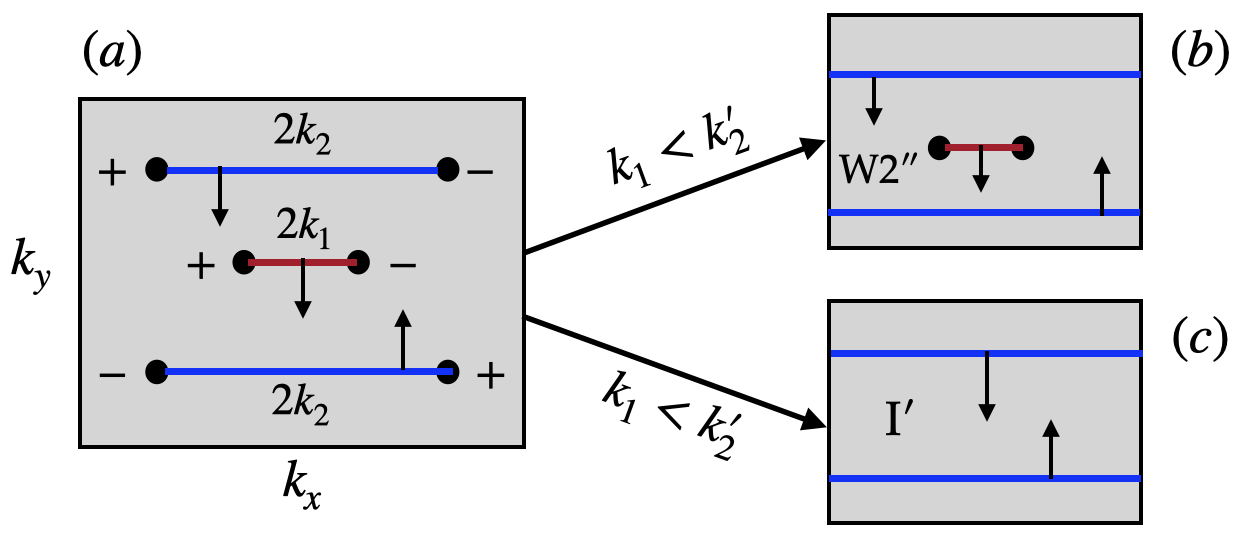} 
\includegraphics[width=0.48\linewidth, height=0.23\linewidth]{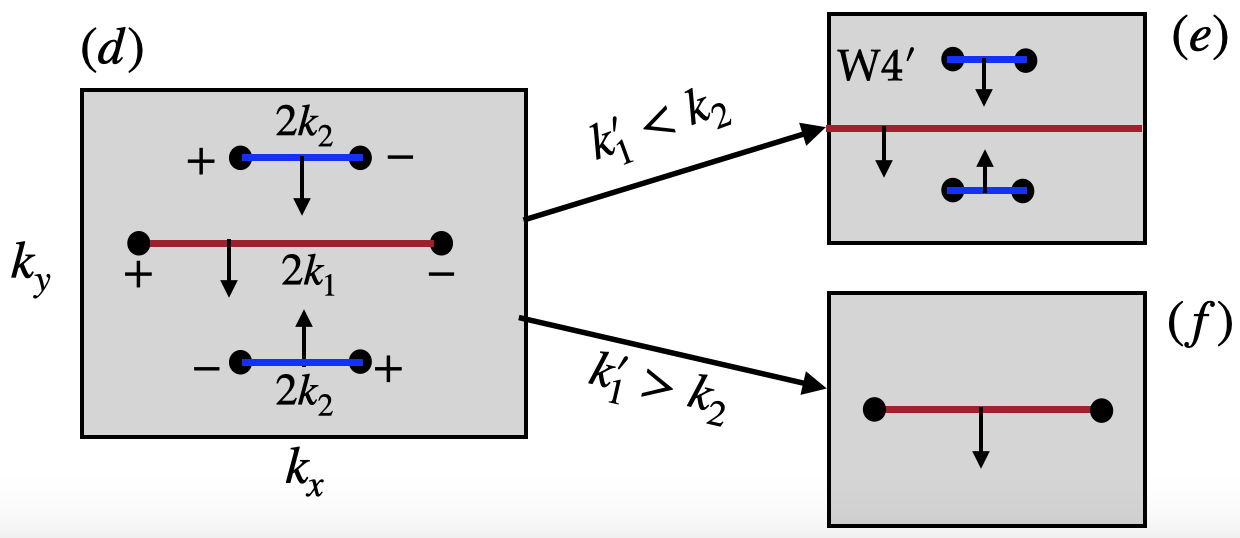} 
\caption{An intuitive representation of pairwise annihilation of WNs  by external magnetic 
field in a WSM with six Weyl nodes. Figures ($a$) and ($d$) depict the projections of 
the WNs (black dots) and the Fermi arcs in the $k_x$-$k_y$ surface BZ.  For a magnetic 
field aligned  in the $y$-direction, only the two separation parameters $k_1$ and  $k_2$  
are relevant. In the first scenario  ($a$), a coexistence phase W2$''$  emerges 
when $k_1<k'_2$ (see ($b$)) and an insulator I$'$ with counter propagating surface 
states  appear when $k_1>k'_2$ (see ($c$))  after pairwise annihilation by magnetic 
fields.  In the second scenario ($d$), pairwise annihilation results in  either a coexistence 
phase W4$'$ or  a WSM with two WNs, depending on the relative values of $k'_1$ and 
$k_2$.  }
\label{fig:WannArc56}
\end{figure*}

The quantity $m_{\alpha}$, $\alpha=0,1,2,...(q-1)$, is equal to $f_1^{\alpha}({\bf k})$ 
i.e. $m_{\alpha} = f_1^{\alpha}({\bf k})$. The Landau level  energy spectrum is 
\begin{align}\label{eq:EBTRP2}
E_n({\bf k}) = \pm \sqrt{ \gamma_n(q, {\bf k}) + \left(f_3({\bf k})\right)^2 }, 
\end{align}
where $\gamma_n(q, {\bf k}) \ge 0$ are the eigenvalues of the positive definite matrix 
${\bf B} {\bf B}^{\dagger}$, and $n=1, 2, 3, ...$ are the Landau level indices. Since 
the spectrum is symmetric about the zero energy, band touching points  are given 
by the zero energy solutions. Clearly,  for  zero energy, we must have  (i) $f_3({\bf k}) 
= 2(\cos{k_2} -\cos{k_y}) = 0$ and  (ii) $\gamma_1 (q, {\bf k})=0$. 
We see from the (i) condition that the band touching along the $k_y$ direction 
remains at  $k_{y0} = \pm k_2$ as we  expected. The corresponding $k_x$ and 
$k_z$ values  at which band touching can occur are determined by the (ii) condition. 
The  condition (ii)  tells that the determinant of the matrix ${\bf B}$ must vanish. Since 
${\bf B}$ is almost an  upper triangular, the determinant can be easily evaluated to be 
\begin{equation}
\begin{aligned}
\textrm{det}({\bf B}) = & \prod_{\alpha} m_{\alpha} - 2^q e^{-i qk_z} \\
 = & 2 \left(T_q(g) - \cos{qk_x} \right)  - 2^q e^{-i qk_z},
\end{aligned}
\end{equation}
where $T_q(g)$ is a Chebyshev polynomial of first kind of degree $q$, and $g=M = 1+\cos{k_1}$.
 Setting  $\textrm{det}({\bf B})=0$ and comparing its real and imaginary parts, we 
arrive at the following two conditions 
\begin{subequations}\label{eq:gapeqTRP2}
\begin{align}
& \sin{qk_z}  = 0 \\
& \cos{qk_x} =  T_q(g) - 2^{q-1}  \cos{qk_z}, 
\end{align}
\end{subequations}
The condition $\sin{qk_z} = 0$ gives two values of 
$k_{z0}=0$ and $\pi/q$ at which gap closing can happen. However, the  solution $k_{z0}=\pi/q$ 
does not satisfy the condition Eq. (\ref{eq:gapeqTRP2}b) because  the right hand side of 
Eq. (\ref{eq:gapeqTRP2}b) is always greater  than the unity for all $q$'s.  Therefore band 
touching along the $k_z$ direction remains at $k_z=0$ and the  corresponding $k_x$ values 
are given by 
\begin{align}\label{eq:gapeqTRP3}
 \cos{qk_x} =  T_q(g) - 2^{q-1}. 
\end{align}
We notice that this condition is identical to the condition in Eq. (\ref{eq:GapConditionTRB}) 
for the time-reversal broken case with two WNs, provided, we have made the  replacement
$k_0 \to k_1$.  The above condition describe a region in the $k_1$ parameter's space for 
gapless solutions. The gapless phase describes  the Weyl semimetal state.  The full phase 
diagram is shown in Fig. \ref{fig:Phaseqy_TRP} for multiple values of $q$.  We notice 
that the phase diagrams are very similar to the phase diagrams of time-reversal broken 
case with two WNs. Now the WSM state has four WNs  and the LCI state is to be replaced 
by the insulator I$'$ which  has a pair of counter propagating Fermi arc surface states 
which are separated by a  distance $k_2$  along the $k_y$-direction in the $k_x$-$k_y$ 
surface BZ. This confirms our prediction derived from the intuitive picture of pairwise annihilation:
The phase  which results in  after  pairwise annihilation by magnetic field aligned along
the $y$-direction is either a normal insulator or an
insulator (I$'$) with counter  propagating  surface states on the $k_x$-$k_y$ surface BZ.

\subsection{Field along $x$-direction}

For magnetic field aligned along the $x$-direction, the separation parameter $k_2$ 
is relevant for pairwise annihilation of Weyl nodes. We do not need to go through 
the whole  calculation to find what would be the  possible  phases. We can easily 
guess the  phase diagram from the intuitive picture of pairwise annihilation of Weyl 
nodes. 
In the zero field  model, the Fermi arcs join projections of WPs which are separated 
along the $k_x$ direction. Since the two Fermi arcs are counter propagating, the 
insulator which results in after pairwise annihilation of WNs either at a point inside 
the BZ  or  at the boundary of the BZ  will be devoid of surface states. Therefore, the 
phase diagram should consist of  of two insulating regions (representing normal 
insulators which have no surface states)  which are separated by a WSM phase in 
the central region. 

\begin{figure*}
\centering
\includegraphics[width=1\linewidth, height=0.32\linewidth]{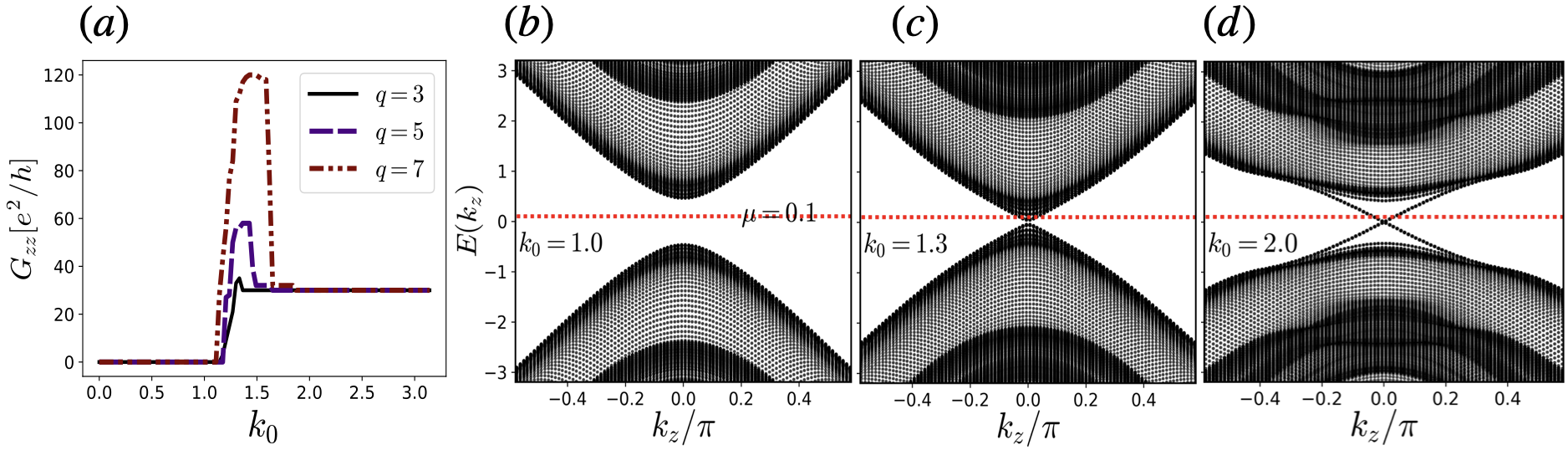} 
\caption{($a$) Longitudinal conductance $G_{zz}$ as a function of separation parameter $k_0$ 
between two  WNs of opposite chirality for three different values of $q = 3, 5, 7$. 
Conductance is computed for a  WSM  slab of length $L_z=100$ and width $L_x=L_y = 25$ 
(the model is defined in Eq. \ref{eq:H0TRB}).  The chemical  potential is fixed  at $\mu=0.1$. 
The conductance $G_{zz}$  is maximum for intermediate separation but  vanishes for small 
separation and drops to finite value for large  separation. The figures (b)-(d) show the 
spectrum of the slab (taken periodic along the transport direction $z$)
for three different values of  $k_0=1.0$ (normal insulator),  $k_0=1.3$ (WSM) and $k_0=2.0$ (LCI) 
for a fixed $q=5$. }
\label{fig:Conductance1TRB}
\end{figure*}

\section{Discussion}
\label{Sec:Discussion}

We have explored the minimal  model of time-reversal broken and time-reversal 
preserved WSM  with two and four WNs  respectively to 
demonstrate how phase diagrams in presence of an external magnetic fields  can be 
derived from an intuitive picture of pairwise annihilation of Weyl nodes.  
As the number of WNs increases, the complexity of solving the model to determine 
the phase diagram grows due to the escalating number of free parameters. The true 
strength of the intuitive representation of the pairwise annihilation process lies in its 
independence from intricate model details. It only necessitates information about the 
locations of WNs and Fermi arc connectivities in the surface  BZ to predict the potential 
phases that may emerge after pairwise annihilation induced by magnetic fields.

Let  us consider a WSM with six WNs and see if there is any new phase which was 
not there in the previous models  with two and four Weyl nodes. Imagine all the WNs are 
located at the $k_x$-$k_y$  plane at $k_z=0$. Suppose the Fermi arcs 
connect the projection of WNs which are separated along the $k_x$ direction as 
depicted in Fig.  \ref{fig:WannArc56}. Assume the magnetic field is applied along the 
$y$ direction so that the separation $k_1$ and $k_2$ (as in Fig.  \ref{fig:WannArc56}) are 
relevant for pairwise annihilation. We have considered two scenarios. In the first
scenario, we have $k_1 \ll \pi/2$ and $k_2 \sim \pi \to k'_2 \ll \pi/2$. Now depending 
on the relative values of $k_1$ and $k'_2$, the pairwise annihilation by magnetic field 
results in either an insulator (I$'$) or  a new coexistent phase W2$''$.  The second 
scenario, where we have $k_1 \sim \pi \to k'_1 \ll \pi/2$ and $k_2 \ll \pi/2$, results in 
either a WSM state with two WNs or a new coexistence phase W4$'$.  So we find that, 
in a WSM  with six WNs, pairwise annihilation of WNs by external fields can lead to 
at least two new  phases which were not possible in a WSM with two and four Weyl nodes.

We have seen that the pairwise annihilation of WNs by external field in a WSM results
in a state which can be an insulator (e.g. NI, LCI, I$'$),  a coexistence phase (e.g. W2$'$, W2$''$, W4$'$)
or a WSM with reduced number of Weyl  nodes. A pertinent question arises: are there any 
experimental signatures of these transitions? One potential quantity to investigate is the 
magnetoconductance. For example, the transition from a WSM state to a normal insulator 
can be distinguished from the transition of a WSM to a LCI state by measuring the 
magnetoconductance. Though,  both the normal insulator and the LCI state  are 
gapped in bulk, the LCI state has protected zero energy surface states. 
Suppose the WNs are at zero energy in the model (as we have in our case). There 
are no states available 
near zero energy in the normal insulating state to carry current. Therefore, we expect 
the conductance, at the transition from the WSM state  to the normal insulating state, to 
drop to zero. However, the conductance at the transition from the WSM to the LCI state 
should be finite because there are finite number of states  near zero  energy due to
the zero energy surface states  in  the LCI state. 

We have computed the (ballistic) magnetoconductance for the WSM model (Eq. \ref{eq:H0TRB}) with 
two Weyl nodes.  The magnetic field is aligned along the $z$-direction. 
The quantity of interest is $G_{zz}$ which measures  the longitudinal conductance 
along the $z$-direction i.e. along the direction of the applied  magnetic field.   
We employ KWANT \cite{Groth_Waintal_2014} simulation  to compute the longitudinal 
conductance $G_{zz}$.  
The conductance $G_{zz}$ is plotted in Fig. \ref{fig:Conductance1TRB} for three different 
$q$ values. Because of  the computational limitation arising due  to the finite  size of the 
system along the transverse directions ($L_x$ and $L_y$), we restrict  ourselves to only 
small $q$ values. The chemical potential is fixed at $\mu=0.1$. We clearly see that the 
conductance vanishes for small WNs separation (normal insulator) and it drops 
but remains finite  for large WNs  separation (LCI state). This demonstrates that  the transition 
from  a WSM  state  to a  normal insulator may be distinguished from the transition of a WSM 
to a LCI state by  measuring the longitudinal conductance in the experiment.



\section{Summary and Conclusion}
\label{Sec:Summary}

An external magnetic field, when aligned in the appropriate direction,  can couple a pair 
of WNs of opposite chirality  and  can potentially annihilate the pair.  Pairwise annihilation 
of WNs occurs when the inverse magnetic length $l^{-1}_B$ becomes close to or larger 
than the momentum space separation $2k_0$ between the two WNs of opposite chirality. 
In this work, we have investigated pairwise annihilation of WNs by external magnetic field
which ranges all the way from small ($l_B \gg a$) to a very large value in the Hofstadter 
regime ($l_B \sim a$). 
We have  shown that pairwise annihilation of WNs by external  magnetic field in a WSM 
with two WNs  results in either a normal insulator or a layered Chern insulator. For a WSM
with more than two WNs which are not collinear, the magnetic field which is applied along 
any of the three  perpendicular directions can induce pairwise annihilation of Weyl nodes. The set 
of phases which  appear for fields along, say, $x$-direction is not identical to the set of 
phases for  fields aligned in the $z$-direction.

We conducted a comprehensive investigation into pairwise annihilation phenomena 
within both the time-reversal broken and time-reversal preserved models of WSMs. 
Our findings reveal that the pairwise annihilation of WNs induced by external magnetic 
fields leads to an emergence of a new state which can be an insulating state 
(e.g., NI, LCI, I$'$), a coexistence phase (e.g., W2$'$, W2$''$, W4$'$), or a WSM with 
a reduced number of Weyl nodes.

We have developed a model independent intuitive representation of pairwise annihilation 
process of WNs  induced by external magnetic fields.  This conceptual framework relies solely on 
information pertaining to the locations of the WNs  and the connectivities of Fermi arcs 
on the surface BZ. With  these essential inputs, our intuitive model accurately predicts
the resulting phases following the pairwise annihilation of WNs induced by external magnetic fields.

This conceptual framework is versatile and can extend its applicability to elucidate the 
pairwise annihilation processes induced by external magnetic fields in other point node 
semimetals, such as three-dimensional Dirac semimetals, as well as two-dimensional 
point node semimetals such as  Weyl semimetals  and Dirac semimetals \cite{Young_Kane_2015, 
Kim_Keun_2017, Jin_Zheng_2020, Feng_Yang_2021, Meng_Zhang_2021, He_Yao_2020, 
You_Su_2019, Abdulla2024protected}. We anticipate further  exploration of these systems 
in the future research.


\begin{acknowledgements}

The author acknowledges the financial support provided by the Infosys Foundation.  
I wish to express my sincere thanks to  ICTS for their warm hospitality during my visit, 
where a significant portion of this work was undertaken. 

\end{acknowledgements}


\appendix

\begin{figure}[t]
\centering
\includegraphics[width=0.49\linewidth, height=0.5\linewidth]{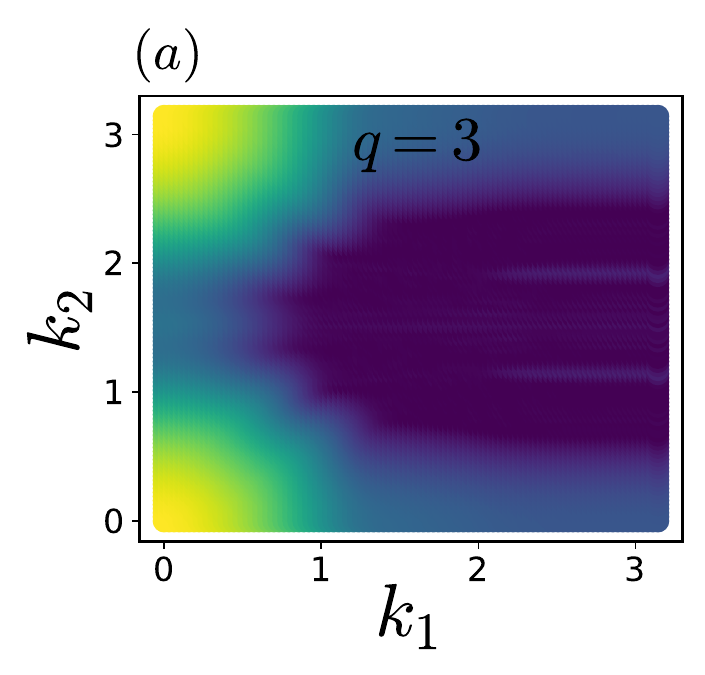} \includegraphics[width=0.49\linewidth, height=0.5\linewidth]{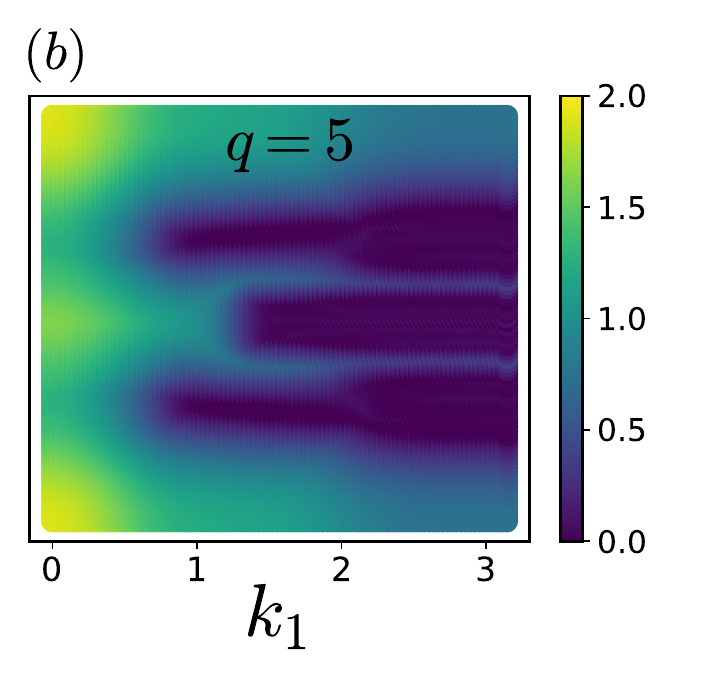} 
\caption{Energy gap of the system (Eq. \ref{eq:HKTRP}) in a slab geometry (finite along the 
$z$-direction) is plotted  as  a function of  the separation parameters $k_1$ and $k_2$.  
The slab has zero energy surface states in the dark-blue regions.  Comparing with the 
phase diagrams Figs. \ref{fig:Phaseqz_TRP}$b$ and   \ref{fig:Phaseqz_TRP}$d$, we 
see that the insulator which is living on the `right insulating region' has zero energy 
surface states. We take a representative point $k_1=2.6$, $k_2=2.2$  from the `right 
insulating  region'   for $q=3$ to show the surface states  in the Fig. \ref{fig:TRP_Surface_Dispersion}.    }
\label{fig:TRP_Surface_Gap}
\end{figure}

\begin{figure}[t]
\centering
\includegraphics[width=0.49\linewidth, height=0.5\linewidth]{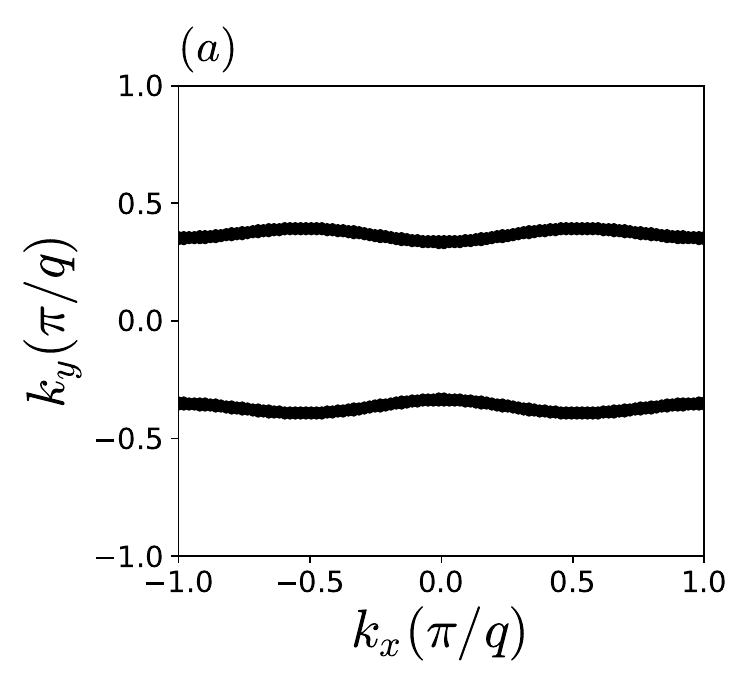} \includegraphics[width=0.49\linewidth, height=0.5\linewidth]{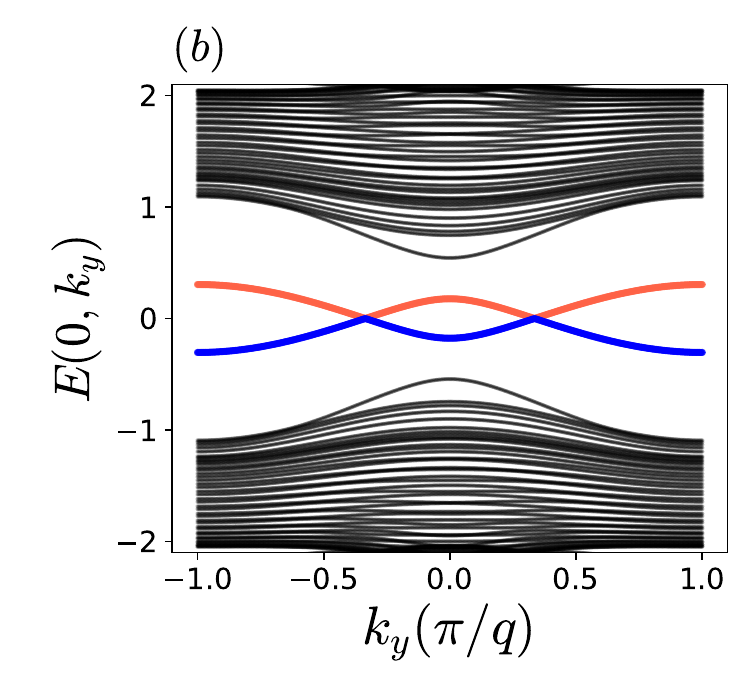} 
\caption{($a$) The zero energy surface states in the $k_x$-$k_y$  surface BZ and ($b$) the 
dispersion  along the $k_y$ direction for $q=3$ of the system (Eq. \ref{eq:HKTRP}) in a slab
geometry (finite along the  $z$-direction).  Values of the separation parameters 
are $k_1=2.6$,  $k_2=2.2$, which represent a model for insulator (see Fig. \ref{fig:Phaseqz_TRP}$b$). 
In figure ($b$), the surface states,  which lie in the bulk  gap of the insulator,  are  highlighted.  }
\label{fig:TRP_Surface_Dispersion}
\end{figure}

\section{The insulating states depicted in the phase diagrams illustrated in Fig. \ref{fig:Phaseqz_TRP}}
\label{App:TRP1}

In Sec. \ref{TRPWSM}, we have studied pairwise annihilation of WNs induced by external fields 
in a WSM with four Weyl nodes. All four WNs are located at the $k_x$-$k_y$ plane 
at $k_z=0$. Pairwise annihilation of WNs by  magnetic field aligned  in the $z$ direction, 
results in a simple phase diagram as shown in Fig. \ref{fig:Phaseqz_TRP}. The phase diagram 
consists of only two phases: a gapless phase (WSM)  and an insulator. 
Let us focus on a phase diagram for a particular value of $q = 100$ (Fig. \ref{fig:Phaseqz_TRP}$l$). 
The entire insulating region in the phase diagram may be split into four subregions: 
left, right, top and bottom insulating regions. As we have argued in main text, the insulators 
which are living in the left, top and bottom regions will not have any surface states.
However the insulator, which is living in the right insulating region where $k'_1 \ll k_1$ and
$k_2 \sim \pi/2$ , can have surface states  (in the $k_x$-$k_y$ surface BZ) in accordance 
with our intuitive picture of pairwise 
annihilation process of Weyl nodes. The bulk of this insulating state is of course trivial and 
the state is adiabatically connected to the adjacent insulating states. We can numerically confirm
whether the insulator living on the right insulating region has  any zero energy surface states. 
Because of  computational limitation, we do this for small values of  $q=3, 5$ (large $q$ values 
require more computational resource). Note that even for small values of $q$, we can split 
the entire insulating regions into for subregions. The previous argument about existence of 
surface states for $q=100$ also applies to the small values of $q$. Therefore, we expect 
the insulator living on the `right insulating region' in the phase diagram for small values of 
$q$ should have zero energy surface states.  We have numerically computed energy gap of  
the system  in a slab geometry (finite in the $z$-direction) to look for the zero energy 
surface states. Since the spectrum of  the Bloch-Hofstadter Hamiltonian  is symmetric 
about the zero energy,  we know for sure that the surface states (if exist) will be at  the zero 
energy.  The energy gap of  the system  in a slab geometry is plotted in Fig. \ref{fig:TRP_Surface_Gap}. 
We can clearly see  the insulator which is living on the `right insulating region' has zero 
energy surface states.  The plots in the Fig. \ref{fig:TRP_Surface_Dispersion} show the zero 
energy surface states in the $k_x$-$k_y$  surface BZ and the dispersion along the $k_y$ direction.


\bibliography{main}

\end{document}